# Réseaux de radio cognitive
Allocation des ressources radio et accès dynamique au spectre

Badr Benmammar et Asma Amraoui

LTT Laboratoire de Télécommunications Tlemcen, UABT, Algérie

# Table des matières

# Introduction

Nous assistons actuellement à la multiplication des normes et des standards de télécommunication vu les progrès récents dans ce domaine. Le nombre croissant de standards normalisés permet d'élargir l'éventail des offres et des services disponibles pour chaque consommateur, d'ailleurs, la plupart des radiofréquences disponibles ont déjà été allouées.

Une étude réalisée par la Federal Communications Commission (FCC) a montré que certaines bandes de fréquences sont partiellement occupées dans des emplacements particuliers et à des moments particuliers. Et c'est pour toutes ces raisons que la Radio Cognitive (RC) est apparue.

L'idée de la RC est de partager le spectre entre un utilisateur dit primaire, et un utilisateur dit secondaire. L'objectif principal de cette gestion du spectre consiste à obtenir un taux maximum de l'exploitation du spectre radio.

Pour que cela fonctionne, l'utilisateur secondaire doit être capable de détecter l'espace blanc, de se configurer pour transmettre, de détecter le retour de l'utilisateur primaire et ensuite cesser de transmettre et chercher un autre espace blanc. Le standard IEEE 802.22, qui est basé sur ce concept, est actuellement en cours de développement.

La RC est une forme de communication sans fil dans laquelle un émetteur/récepteur est capable de détecter intelligemment les canaux de communication qui sont en cours d'utilisation et ceux qui ne le sont pas, et peut se déplacer vers les canaux inutilisés. Ceci permet d'optimiser l'utilisation des fréquences radio disponibles du spectre tout en minimisant les interférences avec d'autres utilisateurs.

Le principe de la RC nécessite une gestion alternative du spectre qui est la suivante : un utilisateur secondaire pourra à tout moment accéder à des bandes de fréquence qu'il trouve libres, c'est-à-dire, non occupées par l'utilisateur primaire possédant une licence sur cette bande. L'utilisateur secondaire devra les céder une fois le service terminé ou une fois qu'un utilisateur primaire aura montré des velléités de connexion.

Les réseaux RC doivent pouvoir coexister pour rendre les systèmes de la RC pratiques, ce qui peut générer des interférences aux autres utilisateurs. Afin de traiter ce problème, l'idée de la coopération entre les utilisateurs pour détecter et partager le spectre sans causer d'interférences est mise en place [AMR12c].

La résolution coopérative de problèmes prend une place prépondérante dans les recherches en IAD (Intelligence Artificielle Distribuée). Un domaine de recherche relativement complexe, dérivé de l'IAD, est celui des Systèmes Multi Agents (SMA). La thématique SMA se focalise sur l'étude des comportements collectifs et sur la répartition de l'intelligence sur des agents plus ou moins autonomes, capables de s'organiser et d'interagir pour résoudre des problèmes.

Nous considérons la coopération comme une attitude adoptée par les agents qui décident de travailler ensemble. Dans le cas de la RC, avant de faire la coopération il faut passer par une autre étape « la négociation », car il y a plusieurs utilisateurs qui veulent satisfaire leurs besoins. La négociation joue un rôle fondamental dans les activités de coopération en permettant aux personnes de résoudre des conflits qui pourraient mettre en péril des comportements coopératifs.

Ce rapport se propose de faire le point sur les diférents aspects qu'ont pu prendre les recherches menées jusqu'à présent sur l'applications des SMA dans le domaine de la RC. Pour cela, nous commençons, dans le premier chapitre par donner un aperçu sur les réseaux sans fils et mobiles, nous parlerons en particulier de la norme IEEE 802.22 qui est une norme de radio cognitive. Le chapitre 2, quant à lui approfondit la notion de RC qui est un domaine technique aux frontières des télécommunications et de l'Intelligence Artificielle (IA). A partir du chapitre 3, le concept agent issu de l'IA s'enrichit aux SMA et aux applications associées. Enfin le chapitre 4 établit un état de l'art sur l'utilisation des techniques d'IA, en particulier les SMA pour l'allocation des ressources radio et l'accès dynamique au spectre dans le domaine de la RC.

# Chapitre 1

# Les réseaux sans fil et mobiles

## 1.1. Introduction

Dans ce chapitre nous nous intéresserons aux réseaux sans fil avec leur fonctionnement, leurs catégories, leurs type… nous nous intéresserons aussi aux réseaux mobiles en présentant les aspects architecturales, les évolutions de la 1ère Génération (1G) jusqu'à la 4ème Génération (4G), nous parlerons aussi de la norme IEEE 802.22 qui est une norme de radio cognitive.

Les termes mobile et sans fil sont souvent utilisés pour décrire les systèmes existants, tels que le GSM (Global System for Mobile communication), IEEE 802.11, Bluetooth, etc. Il est cependant important de distinguer les deux catégories de réseaux que recoupent les concepts de mobile et de sans fil de façon à éviter toute confusion.

Les réseaux mobiles et sans fil ont connu un essor sans précédent ces dernières années. Il s'agit d'une part des réseaux locaux sans fil qui sont rentrés dans la vie quotidienne au travers de standards phares tels que WiFi, Bluetooth, etc. D'autre part du déploiement de plusieurs générations successives de réseaux de télécommunications essentiellement dédiés à la téléphonie (2G, GSM) puis plus orientés vers le multimédia (3G, UMTS : Universal Mobile Telecommunications System).

Aujourd'hui, la majorité des ordinateurs et la quasi-totalité des appareils « mobiles » (tels que les téléphones portables, agendas électroniques, etc.) disposent de moyens de connexion à un ou plusieurs types de réseaux sans fil comme le Wifi, le Bluetooth ou l'Infrarouge. Ainsi, il est très facile de créer en quelques minutes un réseau « sans fil » permettant à tous ces appareils de communiquer entre eux. La difficulté de mise en œuvre tient de la zone de réception, liée à la puissance de l'émetteur, à la détection du récepteur et de la sécurité des données transmises.

L'avantage essentiel que représentent les systèmes de communication est la mobilité. Cet aspect a séduit une grande quantité de personnes. Le service de mobilité permet aux usagers de communiquer sur une zone plus ou moins étendue et de pouvoir poursuivre une communication tout en se déplaçant avec toutefois, des limites en vitesse et en distance. Le système permettant d'offrir ce service au sens le plus large est le système cellulaire, en particulier le système tel que le GSM, dont la couverture peut s'étendre sur des pays voire des continents.

Les évolutions se poursuivent de toute part, tant dans le monde des réseaux spécialisés (capteurs, étiquettes intelligentes, etc.) que des réseaux télécoms. Ceux-ci voient désormais des solutions concurrentes apparaître provenant de divers horizons : monde télécoms classiques, monde des réseaux sans fil avec le WiMAX (Worldwide Interoperability for Microwave Access) voire le monde de la diffusion télévision terrestre et satellite.

Enfin, des réseaux d'une étendue encore plus grande sont en cours de développement sous la norme IEEE 802.22 ou WRAN (Wireless Regional Access Networks). Elle concerne la définition d'une interface d'accès point à multipoint fonctionnant dans la bande de diffusion VHF/UHF-TV. Cette norme doit permettre l'utilisation de ces bandes sans interférer avec les canaux de télévision en activité. Cette solution devrait permettre de couvrir le monde rural avec des accès large bande.

**1.2. Réseaux sans fil**

*1.2.1. Définition*

Un réseau sans fil (en anglais Wireless network) est, comme son nom l'indique, un réseau dans lequel au moins deux terminaux peuvent communiquer sans liaison filaire.

Grâce aux réseaux sans fil, un utilisateur a la possibilité de rester connecté tout en se déplaçant dans un périmètre géographique plus ou moins étendu, c'est la raison pour laquelle on entend parfois parler de "mobilité".

Un réseau local sans fil véhicule les informations soit par infrarouge, soit par onde radio (utilisant généralement la bande de fréquence 2.4 GHz). La transmission par onde radio est la méthode la plus répandue en raison de sa plus large couverture géographique et de son débit plus grand.

Les réseaux sans fil permettent de relier très facilement des équipements distants d'une dizaine de mètres à quelques kilomètres. De plus l'installation de tels réseaux ne demande pas de lourds aménagements des infrastructures existantes comme c'est le cas avec les réseaux filaires, ce qui a valu un développement rapide de ce type de technologies.

En contrepartie se pose le problème de la réglementation relative aux transmissions radioélectriques. En effet, les transmissions radioélectriques servent pour un grand nombre d'applications, mais sont sensibles aux interférences, c'est la raison pour laquelle une réglementation est nécessaire dans chaque pays afin de définir les plages de fréquence et les puissances auxquelles il est possible d'émettre pour chaque catégorie d'utilisation.

Il y a quelques règles simples qui peuvent être très utiles pour concevoir un réseau sans fil:

- Plus la longueur d'onde est grande, plus loin celle-ci ira.

- Plus la longueur d'onde est grande, mieux celle-ci voyagera à travers et autour des choses.

- À plus courte longueur d'onde, plus de données pourront être transportées.

### 1.2.2. Fonctionnement d'un réseau sans fil

Le téléphone sans fil communique avec un correspondant par l'intermédiaire du socle qui fait office de point d'accès (AP) vers le réseau téléphonique.

De même, chaque ordinateur du réseau sans fil muni d'une carte réseau adéquate peut émettre (et recevoir) des données vers (et depuis) un point d'accès réseau. Ce dernier peut être physiquement connecté au réseau câblé et fait alors office de point d'accès vers le réseau câblé.

Bien entendu plus on s'éloigne du point d'accès, plus le débit diminue : pour un débit de 1 Mbps, la portée est de 460 m dans un environnement sans obstacle et de 90 m dans un environnement de bureau classique.

Suivant la manière de communication entres les mobiles, le réseau sans fil offre deux modes de fonctionnement différents : Le mode avec infrastructure et le mode sans infrastructure.

*1.2.2.1. Le réseau avec infrastructure*

En mode avec infrastructure, également appelé le mode BSS (Basic Service Set) certains sites fixes, appelés stations support mobile (Mobile Support Station) ou station de base (SB) sont munis d'une interface de communication sans fil pour la communication directe avec des sites ou unités mobiles (UM), localisés dans une zone géographique limitée, appelée cellule.

A chaque station de base correspond une cellule à partir de laquelle des unités mobiles peuvent émettre et recevoir des messages. Alors que les sites fixes sont interconnectés entre eux à travers un réseau de communication filaire, généralement fiable et d'un débit élevé. Les liaisons sans fil ont une bande passante limitée qui réduit sévèrement le volume des informations échangées. Dans ce modèle, une unité mobile ne peut être, à un instant donné, directement connectée qu'à une seule station de base.

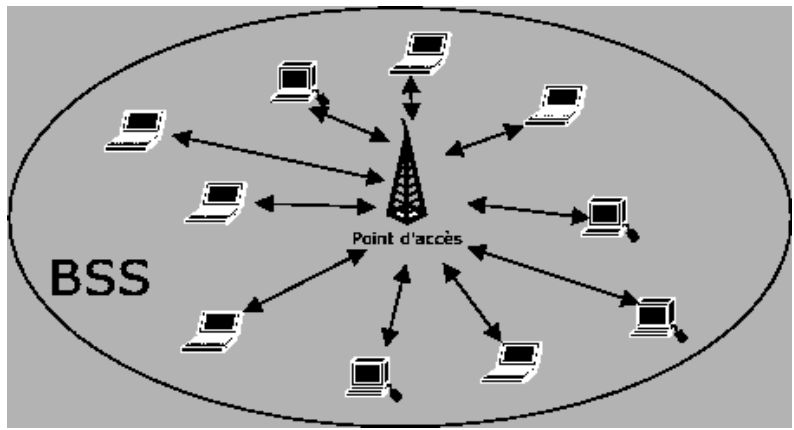

**Figure 1.1.** *Un Basic Service Set*

*1.2.2.2. Le réseau sans infrastructure*

Le réseau sans infrastructure également appelé réseau Ad-hoc ou IBSS (Independent Basic Service Set) ne comporte pas l'entité « site fixe », tous les sites du réseau sont mobiles et communiquent d'une manière directe en utilisant leurs interfaces de communication sans fil. L'absence de l'infrastructure ou du réseau filaire composé des stations de base, oblige les unités mobiles à se comporter comme des routeurs qui participent à la découverte et la maintenance des chemins pour les autres hôtes du réseau.

Ce mode ne bénéficie d'aucune infrastructure. Un groupe de travail appartenant à l'Internet Engineering Task Force (IETF) étudiant ce type de réseau, le définit de la manière suivante :

Un réseau ad-hoc comprend des plates formes mobiles (par exemple routeurs interconnectant différents hôtes et équipements sans fil) appelées nœuds qui sont libres de se déplacer sans contrainte. Un réseau ad-hoc est donc un système autonome de nœuds mobiles. Ce système peut fonctionner d'une manière isolée ou s'interfacer à des réseaux fixes au travers de passerelles. Dans ce dernier cas, un réseau ad-hoc est un réseau d'extrémité.

Ce mode permet de déployer, rapidement et n'importe où, un réseau sans fil. Le fait de ne pas avoir besoin d'infrastructure, autre que les stations et leurs interfaces, permet d'avoir des nœuds mobiles.

La différence entre le mode ad-hoc et le mode infrastructure est que dans le second, toutes les communications passent par l'AP, alors que dans le premier mode la communication entre deux machines se fait directement si elles se trouvent à la portée l'une de l'autre.

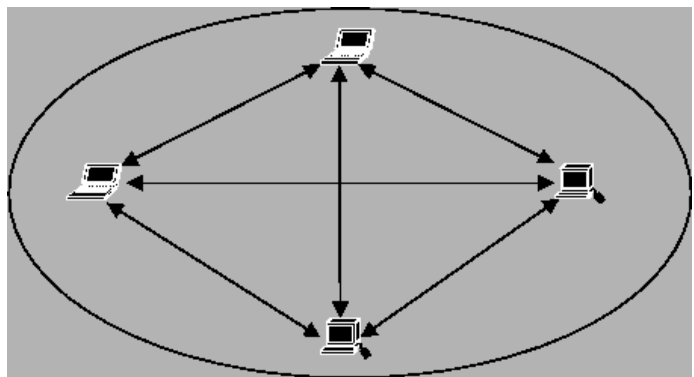

**Figure 1.2.** *La topologie Ad-Hoc*

### *1.2.3. Les catégories de réseaux sans fil*

On distingue habituellement plusieurs catégories de réseaux sans fil, selon le périmètre géographique offrant une connectivité (appelé zone de couverture) :

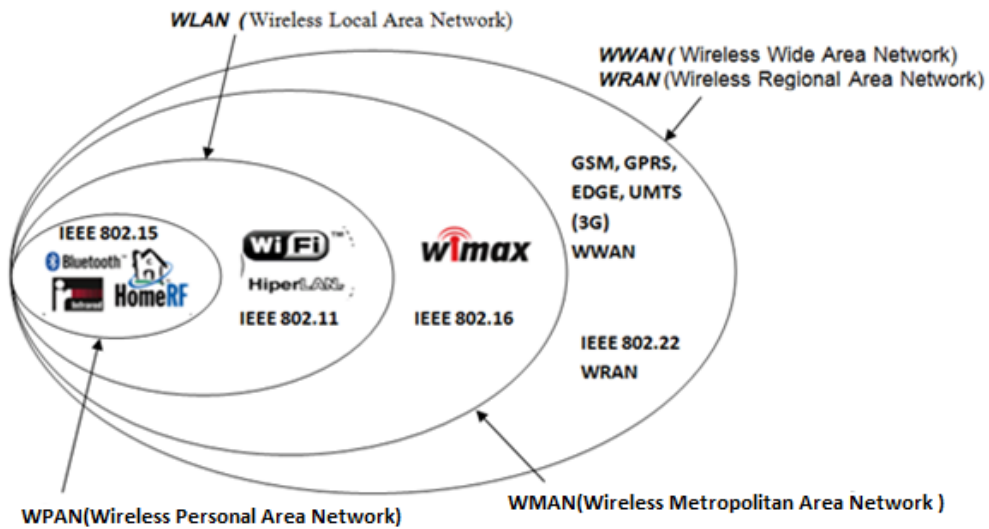

**Figure 1.3.** *Catégories des réseaux sans fil*

#### *1.2.3.1. WPAN (Wireless Personal Area Network)*

Le réseau personnel sans fil est constitué de connexions entre des appareils distants de seulement quelques mètres comme dans un bureau ou une maison.
- Bluetooth
- Home RF

#### *1.2.3.2. WLAN (Wireless Local Area Network)*

Le réseau local sans fil correspond au périmètre d'un réseau local installé dans une entreprise, dans un foyer ou encore dans un espace public. Tous les terminaux situés dans la zone de couverture du WLAN peuvent s'y connecter. Plusieurs WLAN peuvent être synchronisés et configurés de telle manière que le fait de traverser plusieurs zones de couverture est pratiquement indécelable pour un utilisateur.

- IEEE 802.11a, 802.11b, 802.11g
- HiperLan

*1.2.3.3. WMAN (Wireless Metropololitan Area Network)*

Le réseau sans fil WMAN utilise le Standard IEEE 802.16, autrement dit WiMAX (World wide Interoperability for Microwave Access), il fournit un accès réseau sans fil à des immeubles connectés par radio à travers une antenne extérieure à des stations centrales reliées au réseau filaire.

*1.2.3.4. WWAN (Wireless Wide Area Network)*

Le réseau sans fil WWAN englobe les réseaux cellulaires tels que le GSM, GPRS, UMTS, et les réseaux satellitaires.
La distance entre les périphériques peut aller jusqu'à 3Km, le coût de la mise en place d'un tel réseau est plus élevé que celui des réseaux cités au paravent.

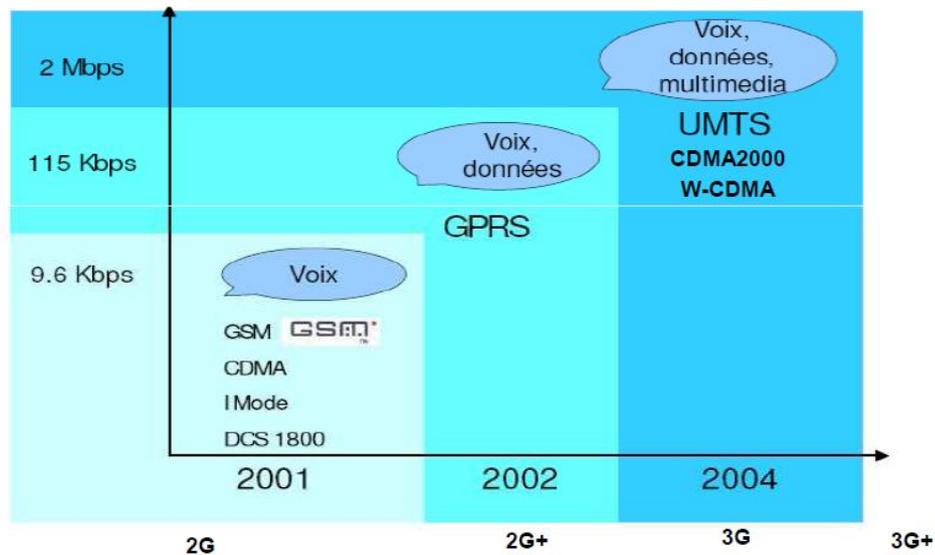

**Figure 1.4.** *WWAN*

*1.2.3.5. WRAN (Wireless Regional Area Network)*

IEEE 802.22 est une norme pour les réseaux régionaux sans fil (WRAN) qui fonctionneront dans des canaux de télévision inutilisés, et fourniront un accès aux services sans fil. La norme finale va supporter des canaux de 6,7 et 8 Mhz pour une opération mondiale. Le WRAN est basé sur l'OFDMA. Cette norme est en cours de développement et est actuellement sous forme d'ébauche.

### *1.2.4. Les différents types de réseaux sans fil existants*

Il existe principalement 2 types de réseaux sans fil :

- Les réseaux utilisant les ondes infrarouges.

- Les réseaux utilisant les ondes radios.

*1.2.4.1. Les réseaux utilisant les ondes infrarouges*

Les ondes infrarouges sont couramment utilisées dans la vie courante (pour les télécommandes de télévisions par exemple). Grâce à elles, on peut créer des petits réseaux, notamment entre des téléphones portables et des ordinateurs.

Le principal inconvénient des réseaux créés avec les ondes infrarouges est qu'ils nécessitent que les appareils soient en face l'un de l'autre, séparés au maximum de quelques dizaines de mètres et qu'aucun obstacle ne sépare l'émetteur du récepteur puisque la liaison entre les appareils est directionnelle. Bien entendu, les seuls réseaux utilisables par cette technologie sont les WPAN.

*1.2.4.2. Les réseaux utilisant les ondes radios*

**Bluetooth**

C'est une spécification industrielle pour les zones de réseaux personnels sans fil (PAN). Il a été lancé par Ericsson en 1994, Ce type de liaison sans fil permet de relier deux appareils via une liaison hertzienne. Ces appareils peuvent être des appareils photo numériques, des PDA, imprimantes.

Il offre des débits moyens (1 Mbits/s en théorie) sur un rayon limité (10 à 30 mètres en pratique).La norme officielle définissant le Bluetooth (dans sa version 1.x) est l'IEEE 802.15.

Sécurisée, cette connexion est transparente uniquement si les deux appareils se connaissent.

Au sein d'un réseau Bluetooth, un appareil sert de maître et jusqu'à 7 périphériques esclaves qui se partagent la bande passante. Il est possible en théorie de faire communiquer jusqu'à 10 groupes d'appareils, soit 80 appareils.

**HomeRF (Home Radio Frequency)**

Soutenu initialement par des acteurs comme Compaq, HP, IBM, Intel et Microsoft, HomeRF a été imaginé avant tout pour un usage domestique. Il utilise les mêmes fréquences que Bluetooth.

En outre, un réseau HomeRF permet de relier des PC portables ou fixes, mais aussi de soutenir des liaisons DECT (Digital Enhanced Cordless Telecommunications), technologie de transport de la voix en mode numérique sur les réseaux sans fil.

HomeRF permet d'adresser 127 nœuds sur un réseau, et 6 liaisons voix simultanées.

**HiperLan (High Performance Lan)**

Elaborée sous la tutelle de l'European Telecommunications Standards Institute, HiperLan est une norme exclusivement européenne. Son but est de créer des environnements flexibles sans fil à haut débit, permettant un fonctionnement ad-hoc. Il dispose d'un code correcteur d'erreur pour obtenir une qualité de transport comparable à celle obtenue dans un réseau local.

**IEEE 802.11**

Avec l'avancement des communications au cours des dernières années, plusieurs technologies visant à répondre aux besoins réels des utilisateurs, la radio a commencé à gagner du terrain depuis l'utilisation des satellites pour un usage personnel, celui-ci était considéré comme une technologie chère, mais, le temps est de plus en plus rentables.

L'IEEE a investi dans l'amélioration de la norme 802.11, avec la même architecture et la technologie, mais avec un débit de données important entre 5 et 11 Mbps, au lieu de pousser la technologie et stimuler les communautés scientifiques et industrielles afin de standardiser, de concevoir et de fabriquer des produits pour ces réseaux. Il existe plusieurs versions de l'IEEE 802.11. En règle générale, plus une version est récente, plus les débits proposés sont élevés.

Plusieurs normes IEEE 802.11 définissent la transmission de données via le medium «hertzien», elles se différencient principalement selon la bande passante, la distance d'émission, ainsi que le débit qu'elles offrent. Les principales extensions sont les suivantes :

**-** *La norme 802.11a*

Cette norme a été développée en 1999 (parfois appelée WiFi5), elle opère dans la bande de fréquence 5 GHz, incompatible avec la fréquence 2,4 GHz. Le schéma de modulation utilisé est OFDM. Ceci limite les interférences et rend possible des vitesses de transmission de données allant jusqu'à 54 Mbps.

Les inconvénients de cette norme sont sa faible portée (15m) et son incompatibilité avec 802.11b.

*- La norme 802.11b, WiFi ou IEEE 802.11hr*

Le terme WiFi, fait référence à cette norme qui fût la première norme des WLAN utilisée par un grand nombre d'utilisateurs, elle a été approuvée en 1999 par l'IEEE. La norme WiFi permet l'interopérabilité entres les différents matériels existants, elle offre des débits de 11 Mbps, avec une portée de 300m dans un environnement dégagé. Elle fonctionne dans la bande de fréquence 2,4GHz, séparée en plusieurs canaux.

*- La norme 802.11b+*

Le 802.11 b+ est dérivé du 802.11b. Il utilise la même gamme de fréquence mais avec des particularités d'un cryptage spécifiques puisque celui-ci se fait sur 64, 128 ou même 256 bits.

Il est tout à fait compatible descendant avec le 802.11b. Un périphérique 802.11b+ acceptera donc la connexion avec les périphériques 802.11b. Par contre, ce standard n'est pas normalisé. Il est donc possible que des appareils 802.11b+ de fabricants différents ne soient pas compatibles.

*- La norme 802.11g*

Cette norme a été développée en 2003. Elle étend la norme 802.11b, en augmentant le débit jusqu'à 54Mbps théorique (30 Mbps réels). Elle fonctionne aussi à 2,4GHz, cette utilisation de la même zone de fréquence devrait permettre de mélanger des points d'accès 802.11b. Le point central adapte sa vitesse en fonction du périphérique connecté, permettant à des clients 802.11b de se connecter.

Grâce à cela, les équipements 802.11b sont utilisables avec les points d'accès 802.11g et vice versa. Cette norme utilise l'authentification WEP statique, elle accepte aussi d'autres types d'authentification WPA (Wireless Protected Access) avec cryptage dynamique (méthode de chiffrement TKIP et AES).

*- La norme  802.11g+*

Cette amélioration du 802.11g est sortie début 2004 et double la vitesse de connexion des 802.11g pour atteindre 108 Mbps en compressant les données. Cette vitesse est donc plus théorique que pratique.

*- La norme 802.11i*

Ratifié en juin 2004, cette norme décrit des mécanismes de sécurité des transmissions. Elle propose un chiffrement des communications pour les transmissions utilisant les technologies 802.11a, 802.11b et 802.11g. La 802.11i agit en interaction avec les normes 802.11b et 802.11g. Le débit théorique est donc inchangé, à savoir 11 Mbps pour la 802.11b et 45 Mbps pour la 802.11g.

*- La norme 802.11e*

Disponible depuis 2005. Elle vise à donner des possibilités en matière de qualité de service (QoS) au niveau de la couche liaison de données, des fonctionnalités de sécurité et d'authentification. Ainsi cette norme a pour but de définir les besoins des différents paquets en termes de bande passante et de délai de transmission de telle manière à permettre notamment une meilleure transmission de la voix et de la vidéo.

*- La norme 802.11n*

Cette norme est seulement normalisée depuis 2009. La vitesse maximum théorique est de 150 à 300 Mbps. Cette vitesse est celle de transport et ne tient pas compte des codes de contrôles, cryptage inclus dans le message. En pratique, le débit effectif devrait être compris entre 100 et 200 Mbps.

Le 802.11n utilise le MIMO (Multiple Input Multiple Output) qui permet d'envoyer et recevoir en utilisant plusieurs antennes simultanément. En modifiant le positionnement des antennes du point d'accès comme de la carte réseau, nous augmentons la distance maximum (mais toujours sous les 100 mètres). Cette solution ne permet pas non plus de "passer les murs" mais permet dans certains cas de les contourner.

Le 802.11n utilise en même temps la bande de fréquences 2,4 GHz et la bande des 5 GHz (utilisée par le 802.11a).

- *WiMAX*

Le WiMAX (World wide Interoperability for Microwave Access) est une connexion sans fil haut débit et longue distance. Elle autorise théoriquement un débit de 70 Mbps sur maximum 50 km, mais en pratique elle offre 10Mbps sur 2 Km.

Basé sur la norme IEEE 802.16, le réseau WiMAX désigne dans le langage courant un ensemble de standards et techniques du monde des réseaux métropolitains sans fil WMAN. WiMAX est principalement fondé sur une topologie en étoile bien que la topologie maillée soit possible.

Il existe différentes versions de WiMAX qui sont utilisées:

- La version 802.16a permet une distance de 20Km maximum avec un débit maximum de 12Mbps. La bande de fréquence utilisée se situe entre 2 et 11GHz. Elle est obsolète.

- La norme 802.16d ou le WiMAX fixe atteint les distances de 50 km. C'est cette norme qui est actuellement commercialisée pour les connexions Internet. Elle n'est nullement conçue pour la mobilité et ne supporte pas le « roaming ».

- La version 802.16e transpose le WiMAX pour la téléphonie mobile avec un taux de transfert de 30 Mbps pour une distance de 3km maximum. Cette solution est en concurrence avec les connexions 3G actuelles (débit de 400 à 700 Kbps). La plage de fréquence se situe entre 2 et 6 GHz.

Le WiMAX est une technologie qui se distingue par deux aspects. D'une part ; le caractère à la fois ouvert, très complet et extrêmement rapide de son processus de normalisation et d'autre part ; le fait d'avoir été le premier à avoir adopté ce qui semble le bon choix en terme de technologie, notamment en matière de modulation, de sécurité et surtout de qualité de service. Le WiMAX utilise le multiplexage OFDM. La figure suivante montre le WiMAX avec ses proches concurrents en termes de couverture géographique et débits offerts.

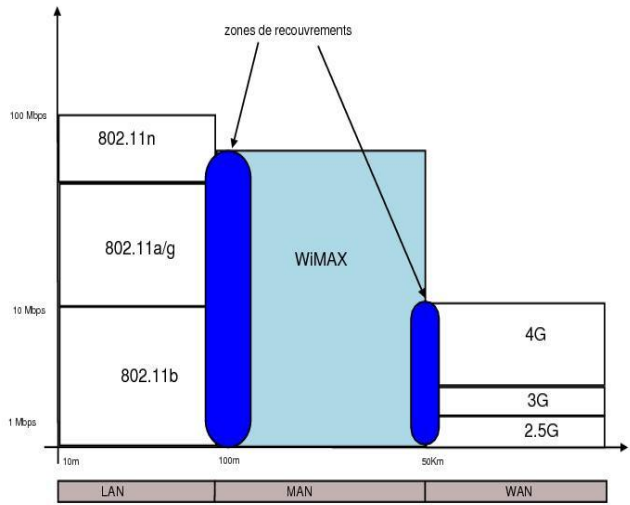

**Figure 1.5.** *Le WiMAX et ses concurrents*

### 1.2.5. Norme IEEE 802.22 :

IEEE 802.22 est un nouveau groupe de travail du comité de normalisation d'IEEE 802 LAN/MAN qui vise à construire l'utilisation sans fil du réseau de région (WRAN) les espaces blancs (canaux qui ne sont pas déjà utilisés) dans le spectre assigné de fréquence de TV.

802.22 indique que le réseau devrait fonctionner dans un point à la base multipoint (P2MP). Le système sera constitué par des stations de base (BS) et équipements de client-lieux (CPE, mentionnés comme points d'Access).

La norme IEEE 802.22 est une norme de radio cognitive visant à doter les régions rurales moins peuplées d'un accès à large bande en utilisant des canaux de télévision vacants. De part le fait que les niveaux du bruit industriel et des réflexions ionosphériques demeurent relativement bas, que les antennes présentent des dimensions raisonnables et que les caractéristiques de propagation sans visibilité directe sont très bonnes, les bandes de radiodiffusion télévisuelle dans la gamme des hautes VHF et des basses UHF se révèlent idéales pour la couverture de vastes régions rurales à faible densité de population.

Le large recours aux technologies de radio cognitive, comme la détection RF, la géolocalisation, l'accès aux bases de données sur les titulaires de station de radiodiffusion, la sélection dynamique de fréquence, vise à assurer la coexistence avec les titulaires de station de radiodiffusion sur une base de non-brouillage ainsi que la coexistence interne avec d'autres systèmes WRAN conformes à la norme 802.22 pour maximiser l'utilisation du spectre.

## 1.3. Réseaux mobiles

### 1.3.1. Sans fil et mobilité

Les termes mobile et sans fil sont souvent utilisés pour décrire les systèmes de communication sans fil existants, il est important de distinguer les deux catégories de réseaux.

Dans les réseaux sans fil, le support de communication utilise l'interface radio : sans cordon, GSM, GPRS, UMTS…

Un utilisateur mobile est défini théoriquement comme un utilisateur capable de communiquer à l'extérieur de son réseau d'abonnement tout en conservant une même adresse.

- Le système sans cordon est un système sans fil mais il n'est pas mobile.

- Certains systèmes tels que le GSM offrent la mobilité et le sans fil simultanément.

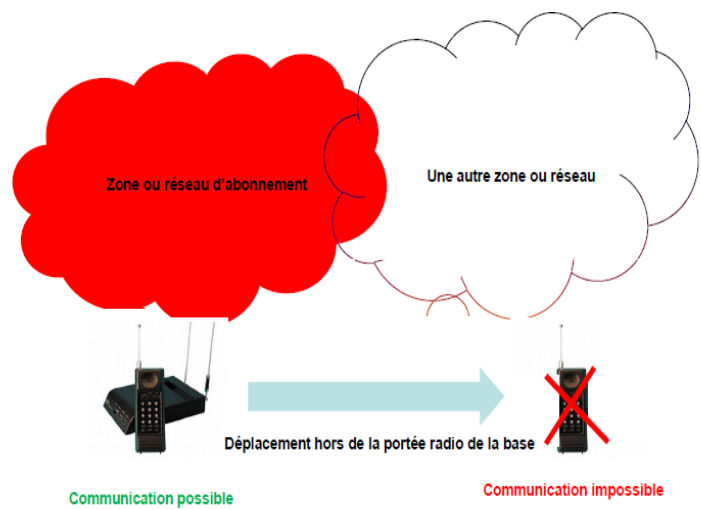

**Figure 1.6.** *Système sans cordon*

*1.3.2 Mobilité*

La mobilité dans les réseaux de communication est définie comme la capacité d'accéder, à partir de n'importe où, à l'ensemble des services disponibles normalement dans un environnement fixe et câblé.

Tandis que l'informatique mobile est définie comme la possibilité pour des usagers munis de périphériques portables ou d'ordinateurs mobiles d'accéder à des services et à des applications évoluées, à travers une infrastructure partagée de réseau, indépendamment de la localisation physique ou du mouvement de ces usagers.

*1.3.3 Architecture cellulaire*

Dans un réseau cellulaire, le territoire couvert ou la zone de couverture desservie est généralement découpée en de petites surfaces géographiquement limitées et communément appelées cellules.

- Picocellule: désigne un espace de desserte de quelques mètres de diamètres.

- Microcellule: réfère à une surface géographique de quelques dizaines de mètres de diamètre.

- Cellule: correspond à une superficie dont le diamètre varie de quelques centaines de mètres à quelques kilomètres.

- Macrocellule: correspond à une étendue géographique de l'ordre de quelques dizaines de kilomètres de diamètre.

- Cellule parapluie: définit une région de quelques centaines de kilomètres de diamètre.

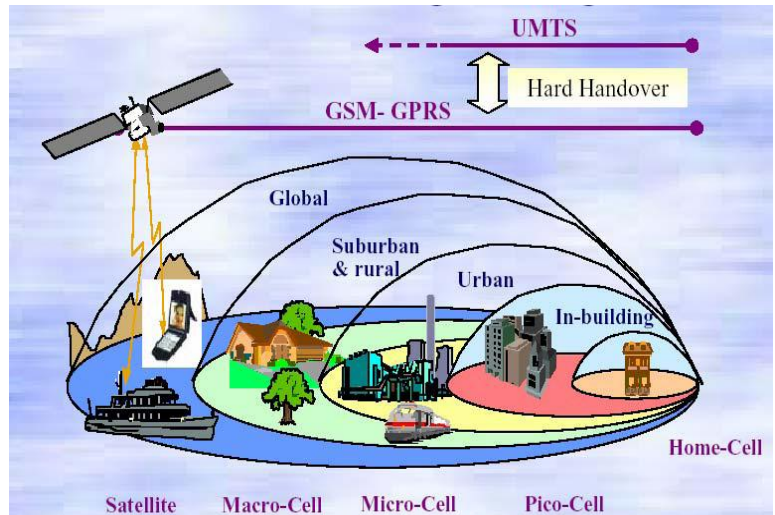
**Figure 1.7.** *Architecture cellulaire*

### *1.3.4. Architecture d'un réseau cellulaire*

Le sous-système radio sert d'interface radio entre chaque unité mobile et le réseau lui-même.

- Station de Base: intègre l'équipement radio/antenne assurant la transmission radio et la signalisation à l'intérieur de cette cellule.

- Contrôleur de stations: gère les ressources radio et les bandes passantes des stations de base associées.

Comme illustré à la figure ci-dessous, les réseaux de communication cellulaires comportent trois niveaux d'hiérarchie. Au premier niveau se trouve le sous-réseau, qui a la charge d'enregistrer le profil d'un abonné. Le deuxième niveau est constitué par la zone de localisation, qui regroupe l'ensemble des cellules, et le dernier niveau par la station de base, qui dessert la cellule.

Si les deux premiers niveaux sont dotés d'intelligence, conformément à la terminologie réseau, la station de base, elle, n'en possède que très peu, assurant un simple rôle de relais radio. Le commutateur, qui gère un ensemble de stations de base, réalise un maximum de procédures pour garantir une connexion : établissement d'appel, gestion du passage intercellulaire, authentification et cryptage, etc.

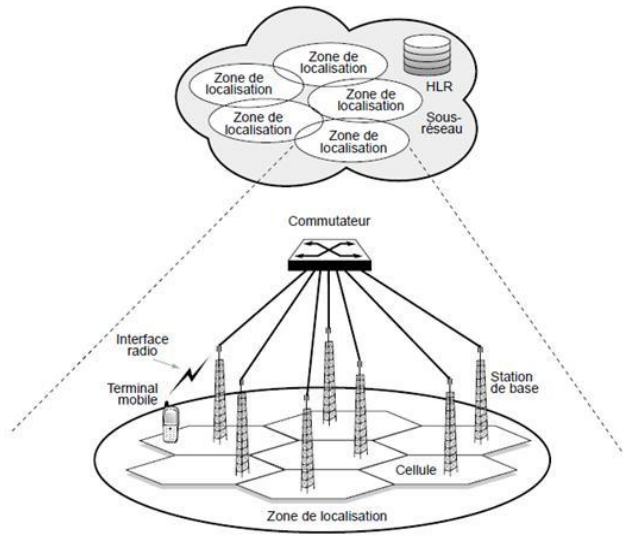

**Figure 1.8.** *Architecture d'un réseau cellulaire*

*1.3.5. Téléphonie*

La téléphonie est un système de télécommunication qui a pour but la transmission de son et en particulier la transmission de la parole. La téléphonie permet des services plus avancés tels que la messagerie vocale, la conférence téléphonique ou les services vocaux.

Un réseau téléphonique est constitué de trois types d'équipements principaux :

- Les terminaux (téléphone, répondeur, modem, fax, serveurs).

- Les systèmes centraux (central téléphonique ou commutateur téléphonique).

- Les liaisons entre différents équipements, tels que les câbles de téléphone (cuivre ou fibre optique) ou les antennes de téléphone mobile.

*1.3.6. Évolution des systèmes cellulaires*

*1.3.6.1. Première génération (1G)*

La première génération de téléphonie mobile possédait un fonctionnement analogique et était constituée d'appareils relativement volumineux.

Problèmes:

- Capacité limitée, car le système est basé sur FDMA.

- Mobilité limitée particulièrement entre réseaux de fournisseurs différents.

- Fraude, absence de mécanismes de sécurité.

- Analogique (canal de contrôle et de la voix).

*1.3.6.2. Deuxième génération (2G)*

**- GSM (Global System for Mobile Communication)**

Ce standard utilise les bandes de fréquences 900 MHz et 1800 MHz en Europe. Aux Etats-Unis par contre, la bande de fréquence utilisée est de 1900 MHz. Ainsi, nous appelons tri-bande, les téléphones portables pouvant fonctionner en Europe et aux Etats-Unis. Il permet de transmettre la voix ainsi que des données numériques de faible volume, par exemple des messages textes SMS (Short Message Service) ou des messages multimédias MMS (Multimedia Message Service).
L'opérateur doit installer des antennes fixes, toutes les antennes définissent une zone de couverture propre à l'opérateur.

Le réseau GSM a pour premier rôle de permettre des communications entre abonnés mobiles et abonnés du réseau téléphonique commuté (RTC réseau fixe). Le réseau GSM s'interface avec le réseau RTC et comprend des commutateurs. La mise en place d'un réseau GSM va permettre à un opérateur de proposer des services de type " Voix " à ses clients en donnant l'accès à la mobilité tout en conservant un interfaçage avec le réseau fixe RTC existant.

**Architecture du réseau GSM**

Dans un réseau GSM, le terminal de l'utilisateur est appelé station mobile. Une station mobile est composée d'une carte SIM (Subscriber Identity Module), permettant d'identifier l'usager de façon unique et d'un terminal mobile, c'est-à-dire l'appareil de l'usager (la plupart du temps un téléphone portable).

Les terminaux (appareils) sont identifiés par un numéro d'identification unique de 15 chiffres appelé IMEI (International Mobile Equipment Identity). Chaque carte SIM possède également un numéro d'identification unique (et secret) appelé IMSI (International Mobile Subscriber Identity). Ce code peut être protégé à l'aide d'une clé de 4 chiffres appelés code PIN.

La carte SIM permet ainsi d'identifier chaque utilisateur, indépendamment du terminal utilisé lors de la communication avec une station de base. La communication entre une station mobile et la station de base se fait par l'intermédiaire d'un lien radio, généralement appelé interface air (ou plus rarement interface Um).

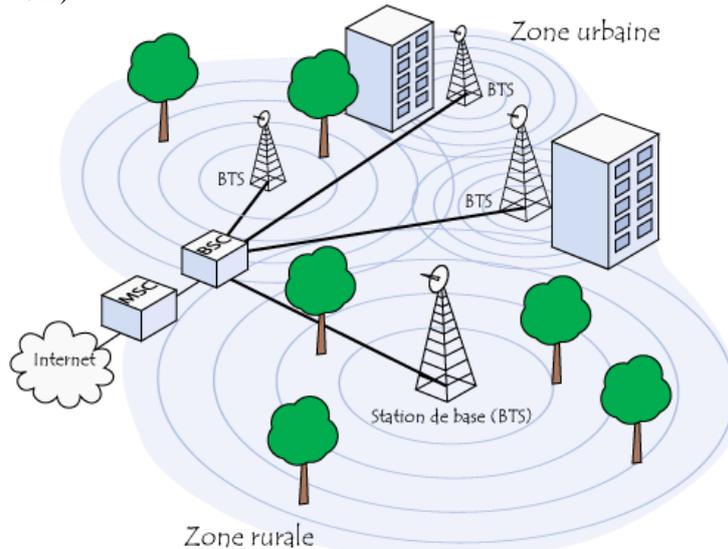

**Figure 1.9.** *Architecture du réseau GSM*

L'ensemble des stations de base d'un réseau cellulaire est relié à un contrôleur de stations, chargé de gérer la répartition des ressources.

L'ensemble constitué par le contrôleur de station et les stations de base connectées constituent le sous-système radio (en anglais BSS pour Base Station Subsystem).

Enfin, les contrôleurs de stations sont eux-mêmes reliés physiquement au centre de commutation du service mobile (en anglais MSC pour Mobile Switching Center), géré par l'opérateur téléphonique, qui les relie au réseau téléphonique public et à internet. Le MSC appartient à un ensemble appelé sous-système réseau (en anglais NSS pour Network Station Subsystem), chargé de gérer les identités des utilisateurs, leur localisation et l'établissement de la communication avec les autres abonnés.

Pour la transmission des données, un protocole WAP (Wireless Application Protocol) a été mis en place qui permet la convergence entre les mobiles et internet, mais GSM ne propose qu'un faible débit (9,6 Kbps), certes satisfaisant pour la voix mais insuffisant pour le transfert des données.

**- GPRS (2.5G)**

Le GPRS (General Packet Radio Service) peut être considéré comme une évolution des réseaux GSM avant leur passage aux systèmes de troisième génération.

Toutefois, la transition du GSM au GPRS demande plus qu'une simple adaptation logicielle. Le GPRS s'inspire des usages devenus courants d'Internet : lors de la consultation de pages Web, une session peut durer plusieurs dizaines de minutes alors que les données ne sont réellement transmises que pendant quelques secondes, lors du téléchargement des pages.

A ce moment, la voix continue de transiter sur le réseau GSM, tandis que les données circulent via le GPRS. Il permet un débit 5x plus élevé que celui du GSM. Il intègre la qualité de service.

**- EDGE (2.75G)**

EDGE (Enhanced Data rates for GSM Evolution) est un réseau de transition entre GPRS et UMTS, il permet un débit encore plus élevé.

EDGE est issu de la constatation que, dans un système cellulaire, tous les mobiles ne disposent pas de la même qualité de transmission. Le contrôle de puissance tente de pallier ces inégalités en imposant aux mobiles favorisés une transmission moins puissante. Cela permet plutôt d'économiser les batteries des terminaux que d'augmenter les capacités de transmission. EDGE permet à ces utilisateurs favorisés de bénéficier de transmissions plus efficaces, augmentant par conséquent le trafic moyen offert dans la cellule.

C'est associé au GPRS qu'EDGE revêt tout son intérêt, notamment grâce au principe d'adaptation de lien. L'adaptation de lien consiste à sélectionner le schéma de modulation et de codage le mieux adapté aux conditions radio rencontrées par le mobile.

*1.3.6.3. Troisième génération (3G)*

Les réseaux 3G ont une grande flexibilité pour introduire de nouveaux services. Les débits sont plus élevés et ils peuvent atteindre les 2Mbps.

**- UMTS (Universal Mobile Telecommunication System)**

Il offre des services de communication sans fil, offre le multimédia en plus de la voix et des données (possibilité de faire une visioconférence, de regarder la télévision), son cout est très élevé.

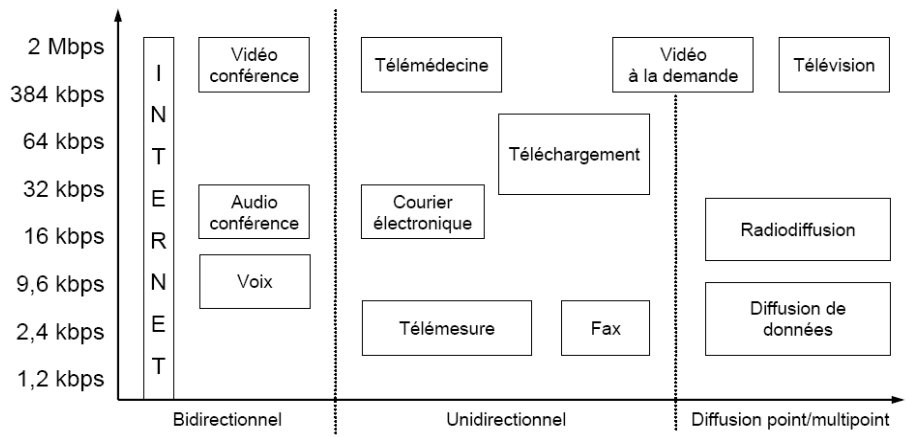

**Figure 1.10.** *Services offerts par le système 3G*

*1.3.6.4. Quatrième génération (4G)*

L'augmentation du nombre d'utilisateurs mobiles du fait du développement de l'internet et de ses applications multimédia, l'apparition rapide des réseaux sans fil et l'évolution de la portabilité des terminaux ont encouragé la mobilité des utilisateurs. Les utilisateurs mobiles ont de plus en plus besoins d'avoir accès à un ensemble riche des services multimédia avancés en utilisant n'importe quel terminal disponible, avec une qualité de service acceptable à travers n'importe quel réseau d'accès disponible. Des débits de données de plus en plus élevés sont utilisés. Les réseaux de 4G utilisent OFDM, OFDMA, ils sont considérés comme étant des réseaux sans coupures surtout avec WLAN et WiMAX.

**1.4. Le WiMAX mobile et la 4G**

L'utilisateur de la 4ème génération de mobile a plusieurs technologies d'accès sans fil à sa disposition. Cet utilisateur veut pouvoir être connecté au mieux, n'importe où, n'importe quand et avec n'importe quel réseau d'accès. Pour cela, les différentes technologies sans fil doivent coexister de manière à ce que la meilleure technologie puisse être retenue en fonction du profil de l'utilisateur et de chaque type d'application et de service qu'il demande. Dans ce contexte, le terminal mobile devra rechercher en permanence le meilleur réseau d'accès en fonction des besoins de l'utilisateur.

Pour la 4ème génération de réseaux mobiles, quelques scénarios potentiels ont été identifiés et les points communs sont montrés ci-dessous :
- Nouveaux équipements de type entrée/sortie seront disponibles pour l'échange rapide de données.
- Nouvelle industrie de semi-conducteur (les terminaux 4G seront disponibles pour tous).
- L'accès aux systèmes mobiles de la quatrième génération sera à bas prix.
- La quantité d'utilisateurs atteindra un niveau élevé.
- Il y aura une grande concurrence entre les applications et les fournisseurs de service pour satisfaire les utilisateurs.
- La qualité de l'accès internet filaire ou sans fil sera égale ou presque identique.
- Les applications multimédias seront utilisées à une grande échelle.
- Les réseaux mobiles devraient être stables, sûrs et disponible tous le temps.
- L'interconnexion devra être facile entre les différents systèmes (GPS, internet, autres réseaux de communication).

Afin de répondre aux différents besoins des utilisateurs, la quatrième génération de mobiles doit satisfaire les conditions techniques suivantes :
- La majorité de personnes peut accéder à la voix ou aux bases de données de services qui sont fournis par les réseaux mobiles (ceci exige une gestion de ressources efficace, par exemple l'utilisation d'une extension ad-hoc dans les systèmes sans fil).
- Le réseau mobile peut être attaché entièrement à internet en raison de son concept de base (de cette façon, la technologie IP serait employée par le réseau mobile (VoIP).
- Le paramètre de la disponibilité de communication dans le réseau doit converger vers 100%.
- Une interface universelle de software/hardware pourrait être normalisée ce qui devrait faciliter le développement de nouveaux services sans aucun problème.

C'est grâce à l'utilisation d'une nouvelle méthode de modulation qui est l'OFDMA avec une nouvelle technologie d'antennes multiples qui est le MIMO que le WiMAX mobile prétend satisfaire les besoins des utilisateurs mobiles.

Le WiMAX mobile peut prétendre concurrencer l'UMTS et constituer la technologie du futur pour une 4ème génération qui n'est pas encore complètement défini. Les réseaux WiMAX mobile devraient représenter en 2012 un quart des équipements de communication mobile au niveau terrestre.

**1.5. Conclusion**

Nous avons présenté le principe de la mobilité des réseaux, leurs applications, la différence entre les réseaux sans fil et les réseaux mobiles, et l'évolution de chaque type de réseau depuis leur apparition.

Le support de la mobilité des réseaux permet de développer l'idée d'une Internet omniprésente, à tout instant, à tout endroit, avec n'importe qui. Les applications multimédia seront les premières à bénéficier de ce type d'environnement.

Les réseaux sans fil en général sont des technologies intéressantes et très utilisées dans de divers domaines comme l'industrie, la santé et le domaine militaire. Cette diversification d'utilisation revient aux différents avantages qu'apportent ces technologies, comme la mobilité, la simplicité d'installation (absence de câblage), la disponibilité (aussi bien commerciale que dans les expériences). Mais la sécurité dans ce domaine reste un sujet très délicat, car depuis l'utilisation de ce type de réseaux plusieurs failles ont été détectées.

Les réseaux sans fil ont connu ces dix dernières années des développements très significatifs qui ont donné aux utilisateurs l'illusion que leurs qualités pouvaient être presque équivalentes à celles des réseaux filaires. Ils ont pris l'habitude que leur mobilité puisse être prise en charge. En revanche, il apparaît également clairement qu'une seule solution technologique, quelle qu'elle soit, ne peut pas répondre à tous les contextes d'utilisation.

Les nouvelles générations de réseaux devront apporter rapidement des solutions exploitables pour offrir la compatibilité entre les différents réseaux de façon à ce que des utilisateurs puissent passer de manière transparente d'un système à un autre.

# Chapitre 2

# La radio cognitive

## 2.1. Introduction

Nous allons étudier dans ce chapitre la radio cognitive dans ses différents aspects : principes, architecture, fonctions et les différents domaines d'application...

Il est aujourd'hui largement reconnu que les systèmes sans fil de communications numériques n'exploitent pas l'intégralité de la bande de fréquence disponible. Les systèmes sans fils de futures générations seront donc amenés à tirer parti de l'existence de telles bandes de fréquence inoccupées, grâce à leur faculté d'écouter et de s'adapter à leur environnement.

En effet, le développement de nouvelles technologies a toujours été dicté par les besoins du moment et la disponibilité de la technique. Nous sommes ainsi passés de la radio analogique à la radio numérique avec tous les progrès qui s'en sont suivis notamment au niveau de la qualité, la rapidité et la fiabilité du transport de l'information mais aussi au niveau de la capacité du réseau.

Avec les années, les besoins se sont amplifiés et de nouvelles solutions techniques sont apparues. Cela a conduit à l'idée de radio logicielle qui au début était prévue pour des applications militaires mais qui s'est progressivement exportée vers le domaine civil. La radio cognitive correspond à l'étape suivante et l'émergence de ce concept est à relier directement avec le besoin de gérer toute cette nouvelle complexité relative à l'environnement du terminal radio.

Certaines bandes et réseaux (GSM, WiFi) sont d'ors et déjà surchargées aux heures de pointe. Pourtant, l'utilisation du spectre n'est pas uniforme : selon les heures de la journée, selon la position géographique, une bande fréquentielle peut être surchargée pendant qu'une autre reste inutilisée. L'idée a donc naturellement émergé de développer des outils permettant de mieux utiliser le spectre.

La radio cognitive est le concept qui permet de répondre à ce défi ; mieux utiliser le spectre, c'est aussi augmenter les débits et rendre plus fiable la couche physique.

**2.2 Radio logicielle (software radio)**

C'est grâce aux travaux de Joseph Mitola que le terme Radio logicielle est apparu en 1991 pour définir une classe de radio reprogrammable et reconfigurable.

La radio logicielle est une radio dans laquelle les fonctions typiques de l'interface radio généralement réalisées en matériel, telles que la fréquence porteuse, la largeur de bande du signal, la modulation et l'accès au réseau sont réalisés sous forme logicielle. La radio logicielle moderne intègre également l'implantation logicielle des procédés de cryptographie, codage correcteur d'erreur, codage source de la voix, de la vidéo ou des données.

Le concept de radio logicielle doit également être considéré comme une manière de rendre les usagers, les fournisseurs de services et les fabricants plus indépendants des normes. Ainsi, avec cette solution, les interfaces radio peuvent, en principe, être adaptées aux besoins d'un service particulier pour un usager particulier dans un environnement donné à un instant donné.

On distingue plusieurs niveaux d'avancement dans le domaine : la radio logicielle est le but ultime intégrant toute les fonctionnalités en logiciel, mais elle impose des phases intermédiaires combinant anciennes et nouvelles techniques, on parle alors de radio logicielle restreinte (software defined radio). Les contraintes de puissance de calcul, de consommation électrique, de coûts, etc. imposent actuellement de passer par cette phase intermédiaire.

*2.2.1. Radio logicielle restreinte (SDR)*

La radio logicielle restreinte est un système de communication radio qui peut s'adapter à n'importe quelle bande de fréquence et recevoir n'importe quelle modulation en utilisant le même matériel [PAL 10].

Les opportunités qu'offre le SDR lui permettent de résoudre des problèmes de la gestion dynamique du spectre. Les équipements SDR peuvent fonctionner dans des réseaux sans fil hétérogènes c'est-à-dire qu'un SDR idéal peut s'adapter automatiquement aux nouvelles fréquences et aux nouvelles modulations.

## 2.3. Radio cognitive

### 2.3.1. Historique

L'idée de la radio cognitive a été présentée officiellement par Joseph Mitola III à un séminaire à KTH, l'Institut royal de technologie, en 1998, publié plus tard dans un article de Mitola et Gerald Q. Maguire, Jr en 1999 [MIT 99]. Connu comme le « Père de la radio logicielle». Dr. Mitola est l'un des auteurs les plus cités dans le domaine. Mitola combine son expérience de la radio logicielle ainsi que sa passion pour l'apprentissage automatique et l'intelligence artificielle pour mettre en place la technologie de la radio cognitive. Et donc d'après lui :

« Une radio cognitive peut connaître, percevoir et apprendre de son environnement puis agir pour simplifier la vie de l'utilisateur »

### 2.3.2. Définition

La cognition regroupe les divers processus mentaux allant de l'analyse perceptive de l'environnement à la commande motrice (en passant par la mémorisation, le raisonnement, les émotions, le langage…).

Le terme radio cognitive est utilisé pour décrire un système ayant la capacité de détecter et de reconnaître son cadre d'utilisation, ceci afin de lui permettre d'ajuster ses paramètres de fonctionnement radio de façon dynamique et autonome et d'apprendre des résultats de ses actions et de son cadre environnemental d'exploitation.

La RC est une forme de communication sans fil dans laquelle un émetteur/récepteur peut détecter intelligemment les canaux de communication qui sont en cours d'utilisation et ceux qui ne le sont pas, et peut se déplacer dans les canaux inutilisés. Ceci permet d'optimiser l'utilisation des fréquences radio disponibles du spectre tout en minimisant les interférences avec d'autres utilisateurs.

La radio cognitive est une nouvelle technologie qui permet, à l'aide d'une radio logicielle, de définir ou de modifier les paramètres de fonctionnement de la fréquence radio d'un nœud réseau (téléphone sans fil ou un point d'accès sans fil), comme par exemple, la gamme de fréquences, le type de modulation ou la puissance de sortie [HAY 05].

Réseaux de radio cognitive : Allocation des ressources radio et accès dynamique au spectre    33Cette capacité permet d'adapter chaque appareil aux conditions spectrales du moment et offre donc aux utilisateurs un accès plus souple, efficace et complet à cette ressource. Cette approche peut améliorer considérablement le débit des données et la portée des liaisons sans augmenter la bande passante ni la puissance de transmissions. La RC offre également une solution équilibrée au problème de l'encombrement du spectre en accordant d'abord l'usage prioritaire au propriétaire du spectre, puis en permettant à d'autres de se servir des portions inutilisées du spectre.

Le SDR Forum (élaboration des normes industrielles du matériel et du logiciel des technologies, en ce moment il mène des travaux de recherches sur la radio cognitive et l'efficacité du spectre) et le groupe de travail P1900 de l'IEEE ont approuvé en Novembre 2007 cette définition :

- "Une radio intelligente est une radio dans laquelle les systèmes de communications sont conscients de leur environnement et état interne, et peuvent prendre des décisions quant à leur mode de fonctionnement radio en se basant sur ces informations et objectifs prédéfinis. Les informations issues de l'environnement peuvent comprendre ou pas des informations de localisation relatives aux systèmes de communication".

Le principe de la radio cognitive, repris dans la norme IEEE 802.22, nécessite une gestion alternative du spectre qui est la suivante : un mobile dit secondaire pourra à tout moment accéder à des bandes de fréquence qu'il juge libre, c'est-à-dire, non occupées par l'utilisateur dit primaire possédant une licence sur cette bande. L'utilisateur secondaire devra les céder une fois le service terminé ou une fois qu'un utilisateur primaire aura montré des velléités de connexion.

On entreprend actuellement de modifier la norme IEEE 802.16 (WiMAX) par le biais de la norme IEEE 802.16h afin de prendre en charge la coexistence et la collaboration dans le même canal. La norme IEEE 802.22, qui vise la coexistence avec les microphones et les systèmes de télévision, présente des caractéristiques semblables. Les deux normes mettent en application les techniques de la radio cognitive.

Un Réseau Cognitif coordonne les transmissions suivant différentes bandes de fréquences et différentes technologies en exploitant les bandes disponibles à un instant donné et à un endroit donné. Il a besoin d'une station de base capable de travailler sur une large gamme de fréquences afin de reconnaître différents signaux présents dans le réseau et se reconfigurer intelligemment.

*2.3.3. Relation entre radio cognitive et radio logicielle restreinte*

L'une des principales caractéristiques de la RC est la capacité d'adaptation où les paramètres de la radio (fréquence porteuse, puissance, modulation, bande passante) peuvent être modifiés en fonction de : l'environnement radio, la situation, les besoins de l'utilisateur, l'état du réseau, la géolocalisation,...etc.

La radio logicielle est capable d'offrir les fonctionnalités de flexibilité, de reconfigurabilité et de portabilité inhérentes à l'aspect d'adaptation de la radio cognitive. Par conséquent, cette dernière doit être mise en œuvre autour d'une radio logicielle. En d'autres termes, la radio logicielle est une "technologie habilitante" pour la radio cognitive.

Bien que de nombreux modèles différents soient possibles, l'un des plus simples modèles conceptuels qui décrit la relation entre la radio cognitive et la radio logicielle restreinte est illustré dans la Figure 2.1.

Dans ce modèle simple, les éléments de la radio cognitive entourent le support radio logicielle restreinte.

Le "cognitive engine" représente la partie chargée de l'optimisation ou du contrôle du module radio logicielle restreinte en se basant sur quelques paramètres d'entrée tels que les informations issues de la perception sensorielle ou de l'apprentissage de l'environnement radio, du contexte utilisateur, et de l'état du réseau.

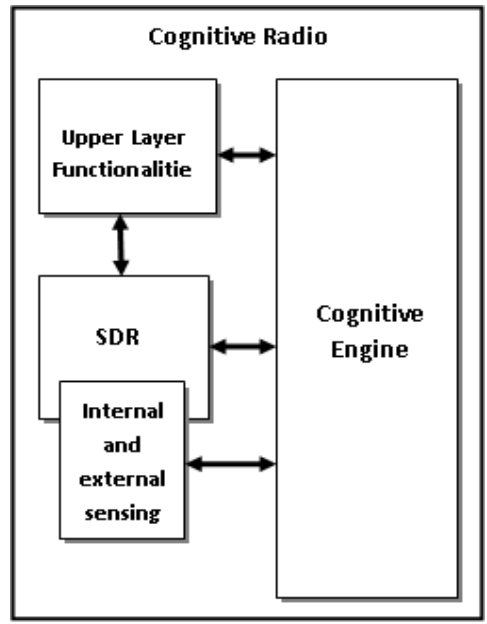

**Figure 2.1.** *Relation entre la radio cognitive et la radio logicielle restreinte*

*2.3.4. Architecture*

Mitola a défini l'architecture d'une radio cognitive par un ensemble cohérent de règles de conception par lequel un ensemble spécifique de composants réalise une série de fonctions de produits et de services.

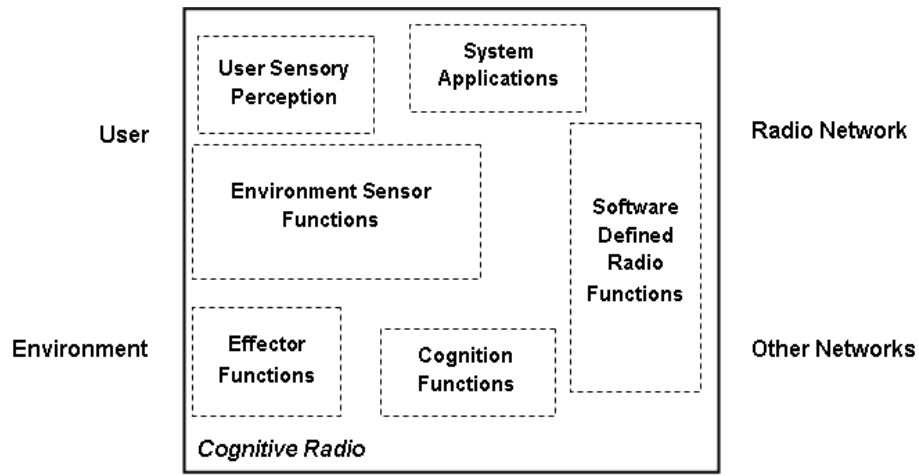

**Figure 2.2.** *Architecture de la radio cognitive*

Les six composantes fonctionnelles de l'architecture d'une radio cognitive sont:

- La perception sensorielle (Sensory Perception : SP) de l'utilisateur qui inclut l'interface haptique (du toucher), acoustique, la vidéo et les fonctions de détection et de la perception.

- Les capteurs de l'environnement local (emplacement, température, accéléromètre, etc.)

- Les applications système (les services médias indépendants comme un jeu en réseau).

- Les fonctions SDR (qui incluent la détection RF et les applications radio de la SDR).

- Les fonctions de la cognition (pour les systèmes de contrôle, de planification, d'apprentissage).

- Les fonctions locales effectrices (synthèse de la parole, du texte, des graphiques et des affiches multimédias).

L'architecture du protocole de la radio cognitive est représentée dans la figure ci-dessous. Dans la couche physique, le RF est mis en œuvre à base de radio définie par logiciel. Les protocoles d'adaptation de la couche MAC, réseau, transport, et applications doivent être conscients des variations de l'environnement radio cognitif. En particulier, les protocoles d'adaptation devraient envisager l'activité du trafic des principaux utilisateurs, les exigences de transmission d'utilisateurs secondaires, et les variations de qualité du canal…

Pour relier tous les modules, un contrôle radio cognitif est utilisé pour établir des interfaces entre l'émetteur/récepteur SDR et les applications et services sans fil. Ce module radio cognitif utilise des algorithmes intelligents pour traiter le signal mesuré à partir de la couche physique, et de recevoir des informations sur les conditions de transmission à partir des applications pour contrôler les paramètres de protocole dans les différentes couches [HOS 09].

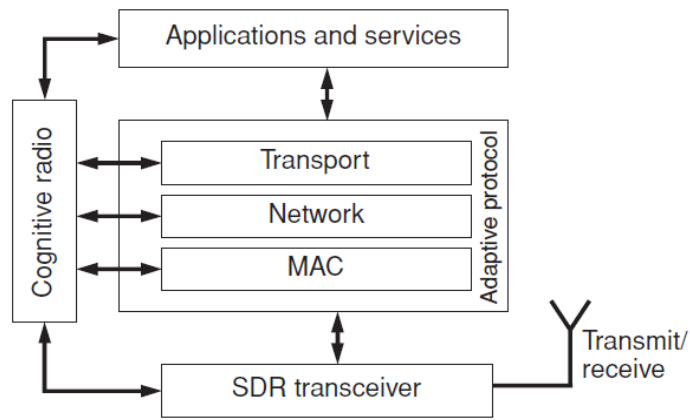

**Figure 2.3.** *Protocoles utilisés par la radio cognitive*

### 2.3.5. Cycle de cognition

La composante cognitive de l'architecture de la radio cognitive comprend une organisation temporelle, des flux d'inférences et des états de contrôle.

Ce cycle synthétise cette composante de manière évidente. Les stimuli entrent dans la radio cognitive comme des interruptions sensorielles envoyées sur le cycle de la cognition pour une réponse. Une telle radio cognitive observe l'environnement, s'oriente, crée des plans, décide, et puis agit [MIT 00].

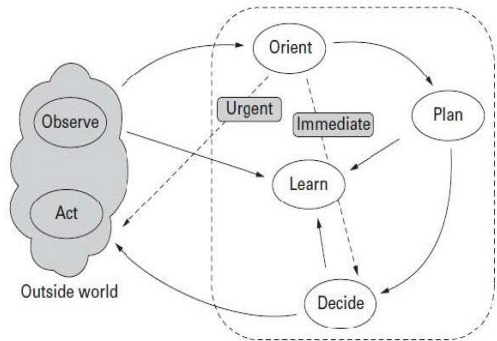 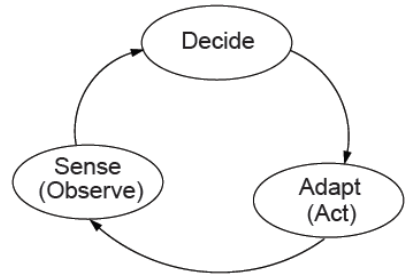

**Figure 2.4.** *Cycle de cognition de Mitola*     **Figure 2.5.** *Cycle de cognition simplifié*

**Phase d'observation (détecter et percevoir)**

La radio cognitive observe son environnement par l'analyse du flux de stimuli entrant. Dans la phase d'observation, la radio cognitive associe l'emplacement, la température, le niveau de lumière des capteurs, et ainsi de suite pour en déduire le contexte de communication. Cette phase lie ces stimuli à des expériences antérieures pour discerner les modèles au fil du temps. La radio cognitive rassemble les expériences en se souvenant de tout.

**Phase d'orientation**

La phase d'orientation détermine l'importance d'une observation en liant à celle-ci une série connue de stimuli. Cette phase fonctionne à l'intérieur des structures de données qui sont analogues à la mémoire à court terme (STM), que les gens emploient pour s'engager dans un dialogue sans forcément se souvenir de tout à la même mesure que dans la mémoire à long terme (LTM). Le milieu naturel fournit la redondance nécessaire pour lancer le transfert de la STM à la LTM. La correspondance entre les stimuli courants et les expériences stockées se fait par reconnaissance des stimuli ou par reliure.

La reconnaissance des stimuli se produit quand il y a une correspondance exacte entre un stimulus courant et une expérience antérieure. La réaction peut être appropriée ou dans l'erreur.

Chaque stimulus est situé dans un contexte plus large, qui inclut d'autres stimuli et les états internes, y compris le temps. Parfois, la phase d'orientation provoque une action qui sera lancée immédiatement comme un comportement réactif « stimulus-réponse ».

Une panne d'électricité, par exemple, peut directement invoquer un acte qui sauve les données (le chemin « immediate » de la phase Action sur la figure). Une perte de signal sur un réseau peut invoquer une réaffectation de ressources. Cela peut être accompli via la voie marquée «urgent» dans la figure.

**Phase de planification**

La plupart des stimuli sont traités avec délibérative plutôt qu'avec réactivité. Un message entrant du réseau serait normalement traité par la génération d'un plan (dans la phase de plan, la voie normale). Le plan devrait également inclure la phase de raisonnement dans le temps. Généralement, les réponses réactives sont préprogrammées ou apprises en étant dit, tandis que d'autres réactions de délibération sont prévues.

**Phase de décision**

La phase de décision sélectionne un plan parmi les plans candidats. La radio peut alerter l'utilisateur d'un message entrant ou reporter l'interruption à plus tard en fonction des niveaux de QoI (Quality of Information) statués dans cette phase.

**Phase d'action**

Cette phase lance les processus sélectionnés qui utilisent les effecteurs sélectionnés qui accèdent au monde extérieur ou aux états internes de la radio cognitive.

L'accès au monde extérieur consiste principalement à composer des messages qui doivent être envoyés dans l'environnement en audio ou exprimés dans différents langages appropriés.

Une action radio cognitive peut également actualiser les modèles internes, par exemple, l'ajout de nouveaux modèles aux modèles internes existants. L'acquisition de connaissances pourrait être achevée par une action qui crée les structures de données appropriées.

**Phase d'apprentissage**

L'apprentissage dépend de la perception, des observations, des décisions et des actions. L'apprentissage initial est réalisé à travers la phase d'observation dans laquelle toutes les perceptions sensorielles sont continuellement comparées à l'ensemble de l'expérience antérieure pour continuellement compter les événements et se souvenir du temps écoulé depuis le dernier événement.

L'apprentissage peut se produire quand un nouveau modèle est créé en réponse à une action. Par exemple, les états internes antérieurs et courants peuvent être comparés avec les attentes pour en apprendre davantage sur l'efficacité d'un mode de communication [NGO 08].

### 2.3.6. Composantes de la radio cognitive

Les différentes composantes d'un émetteur/récepteur radio cognitive qui mettent en œuvre ces fonctionnalités sont présentées dans la figure ci-dessous [HOS 09].

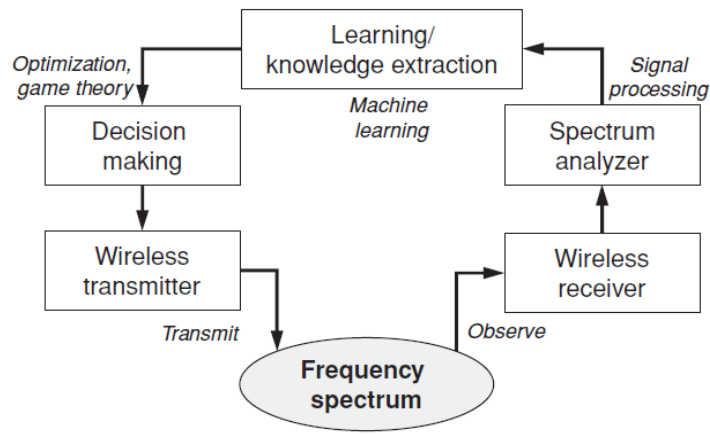

**Figure 2.6.** *Composantes de la radio cognitive*

- Emetteur / Récepteur: un émetteur/récepteur SDR sans fil est le composant majeur avec les fonctions du signal de transmission de données et de réception. En outre, un récepteur sans fil est également utilisé pour observer l'activité sur le spectre de fréquence (spectre de détection). Les paramètres émetteur/récepteur dans le nœud de la radio cognitive peuvent être modifiés dynamiquement comme dicté par les protocoles de couche supérieure.

-Analyseur de spectre (Spectrum analyser): L'analyseur de spectre utilise les signaux mesurés pour analyser l'utilisation du spectre (par exemple pour détecter la signature d'un signal provenant d'un utilisateur primaire et trouver les espaces blancs du spectre pour les utilisateurs secondaires). L'analyseur de spectre doit s'assurer que la transmission d'un utilisateur primaire n'est pas perturbée si un utilisateur secondaire décide d'accéder au spectre. Dans ce cas, diverses techniques de traitement du signal peuvent être utilisées pour obtenir des informations sur l'utilisation du spectre.

   - Extraction de connaissances et apprentissage (Knowledge extraction/learning): L'apprentissage et l'extraction de connaissances utilisent les informations sur l'utilisation du spectre pour comprendre l'environnement ambiant RF (par exemple le comportement des utilisateurs sous licence). Une base de connaissances de l'environnement d'accès au spectre est construite et entretenue, qui est ensuite utilisée pour optimiser et adapter les paramètres de transmission pour atteindre l'objectif désiré sous diverses contraintes. Les algorithmes d'apprentissage peuvent être appliqués pour l'apprentissage et l'extraction de connaissances.

**- Prise de décision (Decision making):** Après que la connaissance de l'utilisation du spectre soit disponible, la décision sur l'accès au spectre doit être faite. La décision optimale dépend du milieu ambiant, elle dépend du comportement coopératif ou compétitif des utilisateurs secondaires. Différentes techniques peuvent être utilisées pour obtenir une solution optimale.
Par exemple, la théorie d'optimisation peut être appliquée lorsque le système peut être modélisé comme une seule entité avec un seul objectif. En revanche, les modèles de la théorie des jeux peuvent être utilisés lorsque le système est composé d'entités multiples, chacun avec son propre objectif. L'optimisation stochastique peut être appliquée lorsque les états du système sont aléatoires.

### *2.3.7. Fonctions de la radio cognitive*

Les principales fonctions de la radio cognitive sont les suivantes:

#### *2.3.7.1. Détection du spectre (Spectrum sensing)*

Détecter le spectre non utilisé et le partager sans interférence avec d'autres utilisateurs. La détection des utilisateurs primaires est la façon la plus efficace pour détecter les espaces blancs du spectre.

L'un des objectifs de la détection du spectre, en particulier pour la détection des interférences, est d'obtenir le statut du spectre (libre /occupé), de sorte que le spectre peut être consulté par un utilisateur secondaires en vertu de la contrainte d'interférence. Le défi réside dans le fait de mesurer l'interférence au niveau du récepteur primaire causée par les transmissions d'utilisateurs secondaires.

*2.3.7.2. Gestion du spectre (Spectrum management)*

Capter les meilleures fréquences disponibles pour répondre aux besoins de communication des utilisateurs.

Les radios cognitives devraient décider de la meilleure bande de spectre pour répondre aux exigences de qualité de service sur toutes les bandes de fréquences disponibles, donc les fonctions de gestion du spectre sont nécessaires pour les radios cognitives. Ces fonctions de gestion peuvent être classées comme suit:

**Analyse du spectre**

Les résultats obtenus de la détection du spectre sont analysés pour estimer la qualité du spectre. Une des questions ici est de savoir comment mesurer la qualité du spectre qui peut être accédée par un utilisateur secondaire. Cette qualité peut être caractérisée par le rapport signal/bruit, la durée moyenne et la corrélation de la disponibilité des espaces blancs du spectre. Les informations sur cette qualité de spectre disponible à un utilisateur par radio cognitive peuvent être imprécises et bruyantes. Des algorithmes d'apprentissage de l'intelligence artificielle sont des techniques qui peuvent être employées par les utilisateurs de la radio cognitive pour l'analyse du spectre.

**Décision sur le spectre**

- ***Modèle de décision*:** un modèle de décision est nécessaire pour l'accès au spectre. La complexité de ce modèle dépend des paramètres considérés lors de l'analyse du spectre.

Le modèle de décision devient plus complexe quand un utilisateur secondaire a des objectifs multiples. Par exemple, un utilisateur secondaire peut avoir l'intention de maximiser son rendement tout en minimisant les perturbations causées à l'usager primaire. Les méthodes d'optimisation stochastique (le processus de décision de Markov) seront un outil intéressant pour modéliser et résoudre le problème d'accès au spectre dans un environnement radio cognitif.

• Compétition / coopération dans un environnement multi utilisateurs : Lorsque plusieurs utilisateurs (à la fois primaires et secondaires) sont dans le système, leur préférence va influer sur la décision du spectre d'accès. Ces utilisateurs peuvent être coopératifs ou non coopératifs dans l'accès au spectre.

Dans un environnement non-coopératif, chaque utilisateur a son propre objectif, tandis que dans un environnement coopératif, tous les utilisateurs peuvent collaborer pour atteindre un seul objectif. Par exemple, plusieurs utilisateurs secondaires peuvent entrer en compétition les uns avec les autres pour accéder au spectre radio (par exemple, O1, O2, O3, O4 dans la figure ci-dessous) de sorte que leur débit individuel soit maximisé. Au cours de cette concurrence entre les utilisateurs secondaires, tous veillent à ce que l'interférence causée à l'utilisateur primaire est maintenue en dessous de la limite de température de brouillage correspondante. La théorie des jeux est l'outil le plus approprié pour obtenir la solution d'équilibre pour le problème du spectre dans un tel scénario.

Dans un environnement coopératif, les radios cognitives coopèrent les unes avec les autres pour prendre une décision pour accéder au spectre et de maximiser une fonction objectif commune en tenant compte des contraintes. Dans un tel scénario, un contrôleur central peut coordonner le spectre de gestion.

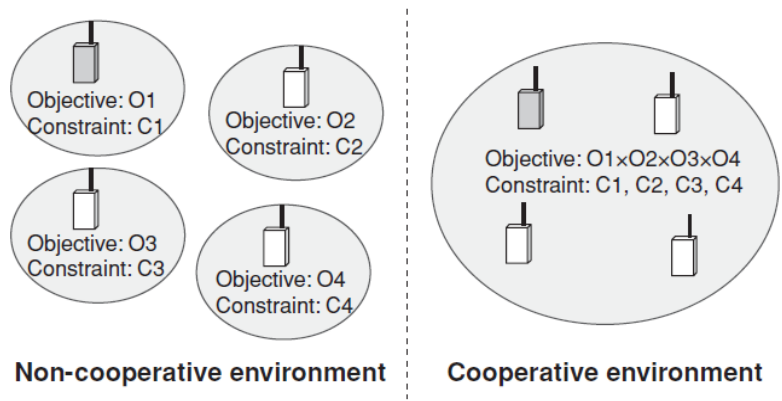

**Figure 2.7.** *Accès au spectre Coopératif et non-coopératif*

- Mise en œuvre distribuée du contrôle d'accès au spectre : Dans un environnement multi utilisateur distribué, pour un accès non-coopératif au spectre, chaque utilisateur peut parvenir à une décision optimale de façon indépendante en observant le comportement (historique / action) des autres utilisateurs du système. Par conséquent, un algorithme distribué est nécessaire pour un utilisateur secondaire pour prendre la décision sur l'accès au spectre de manière autonome.

*2.3.7.3. Mobilité du spectre (Spectrum mobility)*

C'est le processus qui permet à l'utilisateur de la radio cognitive de changer sa fréquence de fonctionnement.

Les réseaux radio cognitifs essayent d'utiliser le spectre de manière dynamique en permettant à des terminaux radio de fonctionner dans la meilleure bande de fréquence disponible, de maintenir les exigences de communication transparentes au cours de la transition à une meilleure fréquence.

- Recherche des meilleures bandes de fréquence

La radio cognitive doit garder une trace des bandes de fréquence disponibles de sorte que si nécessaire (par exemple, un utilisateur autorisé est détecté), il peut passer immédiatement à d'autres bandes de fréquences. Lors de la transmission par un utilisateur secondaire, l'état de la bande de fréquences doit être respecté.

- Auto-coexistence et synchronisation

Quand un utilisateur secondaire effectue un transfert du spectre, deux questions doivent être prises en compte. Le canal cible ne doit pas être actuellement utilisé par un autre utilisateur secondaire (l'exigence d'auto-coexistence), et le récepteur de la liaison secondaire correspondant doit être informé de la non-intervention du spectre (la demande de synchronisation) [HOS 09].

**2.4. Langages de la radio cognitive**

Deux problèmes surgissent. D'abord, le réseau n'a aucun langage standard avec lequel il peut poser ses questions. En second lieu, la destination possède la réponse, mais elle ne peut pas accéder à cette information. Elle n'a aucune description de sa propre structure.

RKRL (Radio Knowledge Representation Language), fournit un langage standard dans lequel de tels échanges de données peuvent être définis dynamiquement. Il est conçu pour être employé par des agents logiciels ayant un haut niveau de compétence conduite en partie par un grand stock de connaissances a priori.

En plus de la langue naturelle, plusieurs langages sont utilisés pour la radio (tableau ci-dessous). L'Union Internationale des Télécommunications (ITU) a adopté les spécifications et le langage de description (SDL) dans ses recommandations. SDL exprime aisément l'état des machines radio, les diagrammes d'ordre de message, et les dictionnaires des données relatifs. L'Institut européen des normes de télécommunications a récemment adopté SDL en tant que l'expression normative des protocoles radio, ainsi on s'attend à ce que la modélisation SDL de la radio continue à avancer. Cependant, SDL manque de primitives pour la connaissance générale des ontologies.

| Langage | Points forts | Points faibles |
|---------|--------------|----------------|
| **SDL** | État des machines, diagramme de séquence, base d'utilisateur très large, connaissances bien codées | Plan de représentation, incertitude |
| **UML** | Ontologies générales, structure, relations | Matériel, propagation RF |
| **IDL** | Interfaces, encapsulation des objets | Informatique générale |
| **KQML** | Primitives (ask/tell), sémantique | Informatique générale |
| **KIF** | Traitement axiomatique des ensembles, relations, frames, ontologies | Informatique générale, matériel, propagation RF |

**Tableau 2.1.** *Langages de la radio cognitive*

Le langage de modélisation unifiée (UML) exprime aisément un logiciel objet, y compris des procédures, des cas d'utilisation, etc. En pratique, il a une présence forte dans la conception et le développement des logiciels, mais il est faible dans la modélisation des dispositifs câblés. En outre, bien qu'UML puisse fournir un cadre de conception pour la propagation radioélectrique, les langages cibles sont susceptibles d'être en C ou en Fortran pour l'efficacité en traçant des dizaines de milliers de rayons d'ondes radio.

Le Common Object Request Broker Architecture (CORBA) définit un langage de définition d'interface (IDL) comme une syntaxe d'exécution indépendante pour décrire des encapsulations d'objets. Ce langage est spécifiquement conçu pour déclarer les encapsulations, il manque de la puissance des langages comme le C ou Java.

Le Knowledge Query and Manipulation Language (KQML), d'autre part, était explicitement conçu pour faciliter l'échange d'une telle connaissance. Basé sur des performatives comme « tell » et « ask ». Le plan de KQML pour prendre un taxi du kiosque de l'information à « Grev Turgatan 16 » emploie la performative Tell pour indiquer le plan du réseau suivant les indications de la figure ci-dessous. Dans cet exemple, la radio avertit également le réseau que son utilisateur compose un certain email et ainsi il va avoir besoin d'une voie de transmission de données de DECT ou de la transmission radioélectrique par paquet de GSM (GPRS) en transit.

```
(Tell: language RKRL: ontology Stockholm/Europe/Global/Universe/Version 0.1
 : Move_Plan (: owner User (: from Kiosk: to "Grev Turgatan 16"):distance 3522m
(: via (Taxi: probability .9) (Foot: probability 0.03))
(: PCS-needs (: DECT 32kbps) (: GSM GPRS) (: backlog Composing-email)))
```

**Figure 2.8.** *Expression d'un plan en KQML*

Le Knowledge Interchange Format (KIF) fournit un cadre axiomatique pour la connaissance générale comprenant des ensembles, des relations, des quantités, des unités, de la géométrie simple, etc. Sa contribution principale est forte. Sa structure est comme celle de LIPS, mais comme IDL et KQML, il n'est pas spécifiquement conçu pour l'usage « interne ».

Le langage naturel souffre des ambiguïtés et de la complexité qui limitent actuellement son utilisation comme langage formel.

La version 0.1 de RKRL a été créée pour remplir ces vides dans la puissance expressive des langages de programmation tout en imposant une parcelle de structure sur l'utilisation du langage naturel [MIT 00].

## 2.5. Domaines d'application de la radio cognitive

Le concept de la radio cognitive peut être appliqué à une variété de scénarios de communication sans fil, nous allons décrire quelques uns :

**- Les réseaux sans fil de prochaine génération** : La radio cognitive devrait être une technologie clé pour la prochaine génération de réseaux sans fil hétérogènes. La radio cognitive fournira des renseignements intelligents à la fois pour l'utilisateur et pour le fournisseur d'équipements. Pour l'utilisateur, un dispositif mobile avec des interfaces d'air multiples (WiFi, WiMAX, cellulaires) peut observer l'état des réseaux d'accès sans fil (la qualité de transmission, débit, délai) et prendre une décision sur la sélection de l'accès au réseau pour communiquer avec. Pour le fournisseur, les ressources radio de plusieurs réseaux peuvent être optimisées pour l'ensemble des utilisateurs de mobiles et de leurs exigences de QoS.

**- Coexistence de différentes technologies sans fil** : Les nouvelles technologies sans fil (IEEE 802.22) sont en cours d'élaboration pour la réutilisation des fréquences radio allouées à d'autres services sans fil (service TV). La radio cognitive est une solution qui fournit la coexistence de ces différentes technologies et services sans fil. Par exemple, IEEE 802.22, basée sur les utilisateurs WRAN peut utiliser efficacement la bande TV quand il n'y a pas d'utilisation du téléviseur à proximité ou quand une station de télévision ne diffuse pas.

**- Services de cyber santé (eHealth services):** Différents types de technologies sans fil sont adoptés dans les services de santé pour améliorer l'efficacité de la prise en charge des patients et la gestion des soins de santé. Cependant, la plupart des dispositifs de soins utilisés sont sans fil et sont limités par les EMI (interférences électromagnétiques) et EMC (compatibilité électromagnétique).

Depuis que les équipements médicaux et les capteurs bio signal sont sensibles aux EMI, la puissance d'émission des appareils sans fil doit être soigneusement contrôlée. En outre, différents dispositifs biomédicaux (équipement et appareils chirurgicaux, de diagnostic et de suivi) utilisent la transmission RF. L'utilisation du spectre de ces dispositifs doit être choisie avec soin pour éviter toute interférence avec l'autre. Dans ce cas, les concepts de la radio cognitive peuvent être appliqués. Par exemple, de nombreux capteurs médicaux sans fil sont conçus pour fonctionner dans les ISM (industriel, scientifique et médicale), et donc ils peuvent utiliser les concepts de la radio cognitive pour choisir les bandes de transmission permettant d'éviter les interférences.

**- Réseaux d'urgence:** les réseaux de sécurité publique et d'urgence peuvent profiter des concepts de la radio cognitive pour fournir la fiabilité et la flexibilité de communication sans fil.

Par exemple, dans un scénario où il y a une catastrophe, l'infrastructure de communication standard peut ne pas être disponible, et par conséquent, un système de communication sans fil adaptatif (soit un réseau d'urgence) peut être nécessaire d'être créé pour soutenir la reprise après sinistre. Ce genre de réseau peut utiliser le concept de la radio cognitive pour permettre la transmission sans fil et la réception sur une large gamme du spectre radio.

**- Réseaux militaires :** Avec la radio cognitive, les paramètres de la communication sans fil peuvent être adaptés de manière dynamique en fonction du temps et de l'emplacement ainsi que de la mission des soldats. Par exemple, si certaines fréquences sont brouillées ou bruyantes, les dispositifs radio cognitifs (émetteurs/récepteurs) peuvent effectuer des recherches pour trouver des bandes de fréquence d'accès de rechange pour la communication.

**2.6. Conclusion**

Nous avons présenté dans ce chapitre des notions importantes concernant la radio cognitive, ainsi que ses principes, en passant par une petite description de la radio logicielle jusqu'aux algorithmes intelligents utilisés dans le domaine de la radio cognitive.

En tenant compte des standards radios existants ou émergents, on peut constater que dans un même environnement, pourraient se trouver, dans une situation de coexistence, différentes interfaces radio (UMTS, GSM/GPRS, WIFI, WIMAX…) qui offrent une variété de services.

D'un point de vue opérateur, une gestion optimisée du spectre s'impose pour pouvoir tirer le maximum de profit de la bande passante globale disponible.

La radio cognitive est un domaine technique aux frontières des télécommunications et de l'intelligence artificielle. Elle est, avant tout, un système radio qui met en place, en plus de sa fonction principale (la communication), un "cycle cognitif" qui lui permet de comprendre son contexte et d'agir en conséquence. Cela offre aux utilisateurs un débit et une QoS accrus, globalement une augmentation du confort dans leurs communications.

Pour assurer ces fonctions, la radio cognitive doit pouvoir déterminer son emplacement géographique, repérer le brouillage, détecter l'occupation du spectre et recueillir de l'information sur la propagation, créant ainsi une sensibilisation à l'environnement radio.

# Chapitre 3

# Systèmes Multi-Agents

## 3.1. Introduction

L'approche classique de l'intelligence artificielle (IA), qui s'appuie sur une centralisation de l'expertise au sein d'un système unique, a montré ses limites dans différents domaines de l'informatique.

L'intelligence artificielle distribuée (IAD) est définie comme étant la branche de l'IA qui s'intéresse à la modélisation du comportement « intelligent » par la coopération entre un ensemble d'agents.

L'IAD propose la distribution de l'expertise sur un groupe d'agents devant être capables de travailler et d'agir dans un environnement commun et résoudre les éventuels conflits.

Actuellement le domaine des systèmes multi-agents est un champ de recherche très actif, qui s'intéresse aux comportements collectifs produits par les interactions de plusieurs agents.

Ce chapitre s'intéresse aux Systèmes Multi-Agents (SMA), et leurs applications dans le domaine des télécommunications ainsi que les interactions, la coopération, la coordination et la communication entre les agents.

**3.2. Définition d'un agent**

Dans la littérature, on trouve plusieurs définitions d'agents qui se ressemblent mais diffèrent selon le type d'application pour lequel l'agent est conçu.

D'après Ferber [FER 95] un agent est une entité physique ou virtuelle qui agit dans un environnement, communique directement avec d'autres agents, possède des ressources propres, est capable de percevoir partiellement son environnement et possède des compétences. En fonction des ressources, des compétences et des communications, un agent tend à satisfaire ses objectifs.

Jennings [JEN 98] a proposé la définition suivante pour un agent : un agent est un système informatique, situé dans un environnement, qui agit d'une façon autonome et flexible pour atteindre les objectifs pour lesquels il a été conçu.

En général, un agent représente un composant logiciel réutilisable qui fournis un accès contrôlé à des services et des ressources.

Le comportement de chaque agent est contraint par des politiques qui sont définies par des agents de contrôle de haut niveau.
La figure suivante représente un agent dans son environnement, l'agent est activé en entrée par les capteurs de l'environnement et produit en sortie des actions.

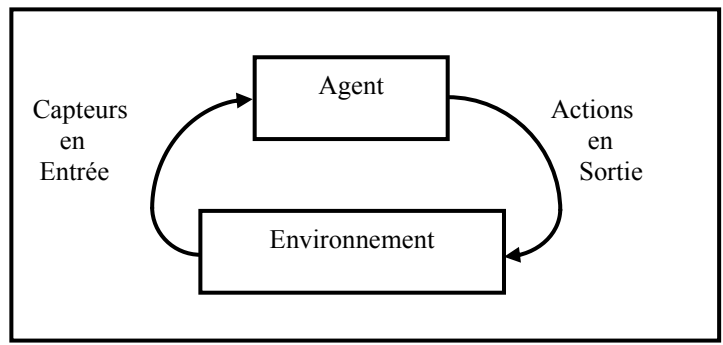

**Figure 3.1.** *L'agent dans son environnement*

*3.2.1. Les caractéristiques multidimensionnelles d'un agent*

Un agent est caractérisé par :

- *La nature* : un agent est une entité physique ou virtuelle.
- *L'autonomie* : un agent est indépendant de l'utilisateur et des autres agents.

- *L'environnement* : c'est l'espace dans lequel va agir l'agent, il peut se réduire au réseau constitué par l'ensemble des agents.
- *La capacité de représentation* : l'agent peut avoir une vision très locale de son environnement mais il peut aussi avoir une représentation plus large de cet environnement et notamment des agents qui l'entourent.
- *La communication* : l'agent aura plus ou moins de capacité à communiquer avec les autres agents.
- *Le raisonnement* : l'agent peut être lié à un système expert ou à d'autres mécanismes de raisonnements plus ou moins complexes.
- *L'anticipation* : l'agent peut plus ou moins avoir les capacités d'anticiper les événements futurs.
- *L'apprentissage* : un agent aura plus ou moins tendance à retirer, stocker et réutiliser des informations extraites ou reçus de son environnement.
- *La contribution* : l'agent participe plus ou moins à la résolution de problèmes ou à l'activité globale du système.
- *L'efficacité* : l'agent doit avoir la rapidité d'exécution et d'intervention.

### *3.2.2. Architecture concrète pour un agent*

Il existe quatre classes d'agents :
- Les agents logiques : les décisions prises par les agents sont basées sur des déductions logiques.
- Les agents réactifs : les décisions prises par les agents sont basées sur une correspondance entre les situations et les actions.
- Les agents BDI : les états internes des agents sont exprimés sous la forme de croyance (Belief), de désirs (Desire) et d'intentions (Intention), la prise de décision est basée sur l'état interne de l'agent.
- Les agents multi-niveaux : l'approche multi-niveaux est utilisée pour organiser les connaissances internes des agents.

*3.2.2.1. Architecture d'agents logiques*

Les connaissances des agents sont décrites sous la forme d'expression logique. L'agent utilise la déduction logique pour résoudre les problèmes et pour caractériser son comportement.

Pour comprendre les principaux problèmes de cette approche, on examine un agent dont la base de faits est constituée des formules logiques suivantes :

Ouvert (valve).

Température (réacteur).

Pression (réservoir).

Les formules représentent l'environnement de l'agent, si l'agent pense que la valve est ouverte, alors il possède le fait Ouvert (valve) dans sa base, mais la présence de ce fait n'implique pas que la réalité de son environnement soit en accord avec ce fait.

Pour cela, il suffit que le capteur fonctionne mal, ou que le raisonnement qui a conduit à la production de ce fait soit faut ou que l'interprétation de la formule Ouvert soit complètement différente selon le concepteur de l'agent et l'agent lui-même.

Dans la pratique, l'approche logique est irréalisable dans des environnements complexes, et fortement dynamiques.

*3.2.2.2. L'architecture réactive*

Les approches réactives sont issues des problèmes et limites rencontrées par l'approche logique.

Pour bien mesurer l'apport des approches purement réactives, nous allons examiner l'architecture de « Subsomption » (de Subsumer, mettre sous) de Rodney Brooks [BRO 86], qui est considérée comme la plus représentative.

Dans cette architecture le comportement de l'agent est vu à travers une fonction « agir » qui lui est propre.

La fonction « agir » décide de l'action à accomplir en fonction des informations sur l'environnement de l'agent.

L'agent possède des fonctions et des tâches spécifiques à accomplir, qui fonctionnent en parallèle, les fonctions manipulent une représentation symbolique simplifiée et ne raisonnent pas directement avec cette représentation.

Les règles de décision sont de la forme situation $\rightarrow$ action, elles font correspondre une action à un ensemble de perceptions.

*3.2.2.3. Architecture BDI*

Dans cette approche le raisonnement nécessite deux processus importants, le premier processus fixe les buts à atteindre c'est-à-dire se demander quoi faire, le deuxième processus pose la question comment faire pour les atteindre [WOO 99].

David Kinny et Michael Georgeff [KIW 91] ont étudié dans ce contexte d'agent BDI, les performances des agents « imprudents » et celles des agents « précautionneux ».

Le paramètre déterminant est le taux de changement du monde dans lequel évoluent ces agents, un taux faible favorise les agents « imprudents » alors qu'un taux élevé favorise les agents « précautionneux ».

Formellement un agent BDI est caractérisé par :

Bel : l'ensemble des croyances possibles.

Des : l'ensemble des désirs possibles.

Inten : l'ensemble des intentions possibles.

L'état de l'agent est décrit à tout moment par le triplet (B,D,I) où $B \subseteq Bel$, $D \subseteq Del$, $I \subseteq Inten$.

Chaque agent BDI est décomposé en au moins sept niveaux fonctionnels, ces niveaux sont les suivants [WOO 99] :

- Un ensemble de croyances courantes (B) sur son environnement.

- Une fonction de révision de ses croyances (brf) qui calcule ses nouvelles croyances à partir des croyances courantes et de ses nouvelles perceptions de son environnement.

- Une fonction de génération de ses options pertinentes (options) qui représentent ses désirs possibles conformément à ses intentions, cette fonction est responsable des actions mises en œuvre, elle doit produire des options consistantes.

- Une fonction de filtre (filtre) qui représente la phase initiale (quoi faire) de son processus de raisonnement, elle active ses nouvelles intentions en fonctions de ses croyances, options, et intentions courantes. Cette fonction élimine les intentions devenues irréalistes ou incohérentes.

- Un ensemble d'intentions courantes (I), représentant ses centres d'intérêts actuels.

- Une fonction de sélection (exécuter) de l'action à exécuter, cette fonction renvoie une intention exécutable qui correspond à une action.

- Une fonction « agir » de décision.

La figure suivante montre l'architecture d'un agent BDI.

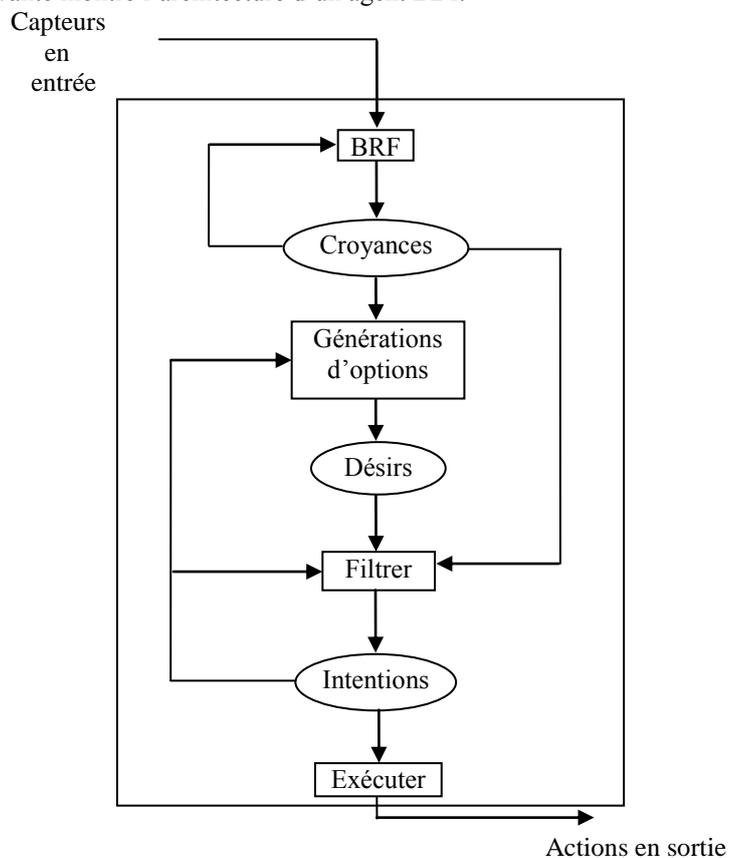

**Figure 3.2**. *Architecture d'un agent BDI*

*3.2.2.4. Architecture multi-niveaux*

L'objectif des architectures multi-niveaux est de faire une synthèse constructive sur les deux approches réactive et pro-active. L'approche pro-active veut dire que l'agent est capable, sur sa propre initiative, de fixer des buts pour atteindre ses objectifs.

Dans ces architectures il existe au moins deux modes de contrôle des échanges d'information entre les niveaux [WOO 99].

- Le contrôle horizontal : les modules sont connectés aux capteurs en entrée et à des actions en sortie, chaque module se comporte comme un agent, l'architecture comporte n niveaux et m actions possibles par niveau, donc nous avons mn interactions possibles.
La figure suivante montre l'architecture horizontale.

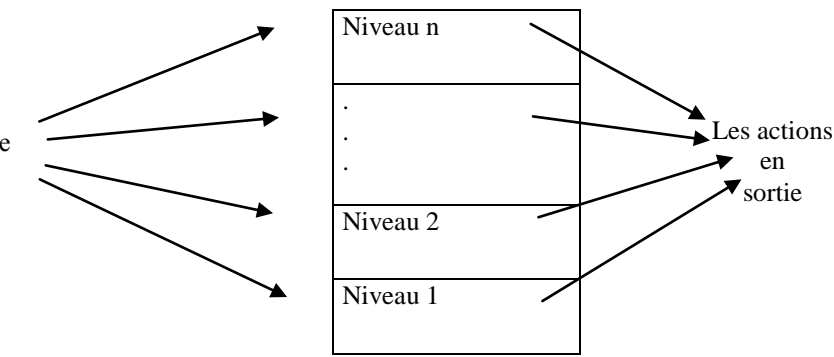

**Figure 3.3.** *Architecture horizontale*

- Le contrôle vertical : il y a un module qui gère les capteurs en entrée et un autre qui gère les actions en sortie. On utilise un contrôle des flux d'information entre les niveaux, le contrôle est basé soit sur le mode à une passe soit sur le mode à deux passes.

Les flux d'information dans le mode à une passe arrivent sur un module spécialisé, puis travers en séquence les autres modules jusqu'au dernier qui pilote la sortie.

Dans le mode à deux passes, les flux d'information suivent le même chemin que dans le mode à une passe, puis redescendent l'architecture en sens inverse pour revenir au module interface de l'agent.

La figure suivante représente l'architecture verticale.

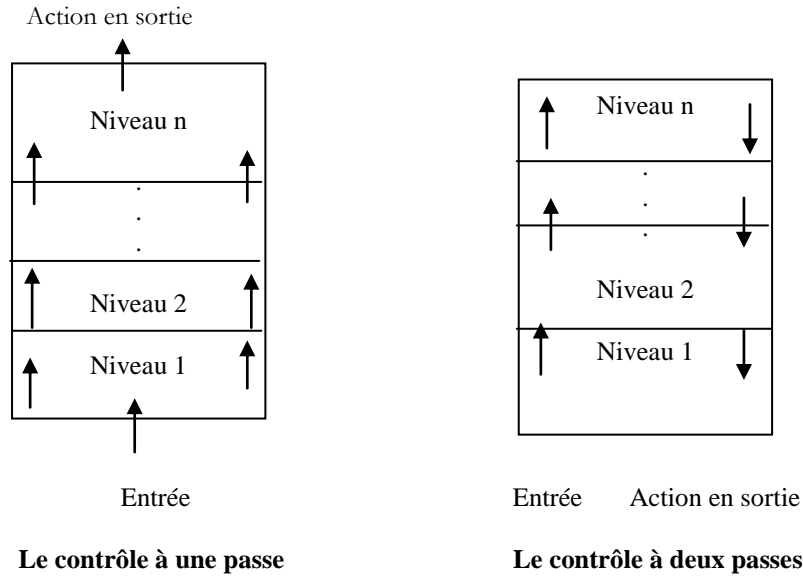

**Le contrôle à une passe**       **Le contrôle à deux passes**

**Figure 3.4.** *Architecture verticale*

### *3.2.3. Modèle type d'un agent*

- Frasson et Gouardères [FRA 96] ont proposé le modèle agent « ACTOR » qui représente une architecture implémentée avec trois niveaux de description (réactif, adaptatif et cognitif).

- Le 1er niveau (réactif) : il permet des comportements réflexifs basés sur une reconnaissance plus ou moins précise de modèles prédéfinis, ce niveau n'a pas de capacité d'apprentissage.

- Le 2ème niveau (adaptatif) : l'apprentissage autorise la modélisation de comportements des agents planificateurs qui sont chargés des stratégies ainsi que des interactions avec les autres agents.

- Le 3ème niveau (cognitif) : il autorise l'apprentissage et la modélisation des modes de raisonnement employés par l'agent cognitif.

La figure suivante représente les trois niveaux de description du le modèle agent « ACTOR ».

| |
|---|
| Le niveau cognitif |
| Le niveau adaptatif |
| Le niveau réactif |

**Figure 3.5.** *Les niveaux de description du modèle agent « ACTOR »*

### 3.3. Les Systèmes Multi-Agents

Un système multi-agents est un ensemble organisé d'agents. Il est constitué d'une ou plusieurs organisations qui structurent les règles de cohabitation et de travail collectif entre agents, dans un même système, un agent peut appartenir à plusieurs organisations [BRI 02].

Les systèmes multi-agents existants sont composés d'agents réactifs ou cognitifs suivant le problème traité.

Dans le cadre de la psychologie cognitive, les systèmes peuvent appartenir à trois grandes catégories [HER 98] :

- Les systèmes multi-experts.

- Les systèmes multi-robots.

- Les systèmes multi-agents de simulation.

- Les systèmes multi-experts : les agents dans les systèmes multi-experts sont virtuels, le système modélise l'interaction de plusieurs agents cognitifs, spécialistes de leur domaine et requis pour l'accomplissement d'une tâche complexe.

- Les systèmes multi-robots : le système est constitué de robots chargés de tâches collectives, le robot est considéré comme un agent artificiel ayant une existence physique.

- Les systèmes multi-agents de simulation : le système est constitué d'agents réactifs et les simulations concernent des modèles biologiques comme la sélection naturelle ou la reproduction.

*3.3.1. Communication entre agents*

Un agent doit être capable de communiquer avec les autres agents. Les agents communiquent entre eux en échangeant des messages. L'envoi et la réception de messages forment le niveau élémentaire de communication entre les agents.

L'agent peut participer à un dialogue en étant passif ou actif. Un agent passif doit accepter les questions des autres agents et répondre à leurs questions. Un agent actif doit proposer et envoyer des interrogations.

Dans un dialogue les agents alternent des rôles actifs et passifs, et échangent des séries de messages en respectant des protocoles biens précis, ce sont les protocoles de coordination, de coopération et de négociation.

*3.3.1.1. Les protocoles de coordination*

Les agents fonctionnent suivant deux principes : les engagements et les protocoles de coordination.

Les engagements sont des structures qui permettent à un agent de s'engager à faire un ensemble d'actions vis-à-vis de lui-même et des autres agents. Les protocoles de coordination lui permettent de gérer ces engagements dans le cas où les circonstances dans lesquelles ils ont été élaborés, évoluent [HUH 99].

Ils définissent aussi sous quelles conditions les engagements peuvent être revus et quelles sont alors les actions à prendre.

Les protocoles de coordination aident les agents à gérer leurs engagements, mais ne disent rien sur ce qu'un agent doit faire vis-à-vis des autres agents, quand l'agent modifie ses engagements.

*3.3.1.2. Les protocoles de coopération*

La coopération entre les agents consiste à décomposer les tâches en sous-tâches puis à les répartir entre les différents agents, il existe plusieurs décompositions possibles, le processus de décomposition doit donc tenir compte des ressources disponibles et des compétences des agents.

La décomposition peut être faite soit par le concepteur, ou par les agents eux-mêmes, grâce à des techniques de planification hiérarchisées ou encore elle est inhérente à la représentation du problème [HUH 99].

Les mécanismes utilisés pour distribuer les sous-tâches aux agents sont :

- L'économie de marché : les tâches sont allouées aux agents sur le principe de l'offre et de la demande, elles sont considérées comme des marchandises qui ont une valeur (achat/vente).

- Le contrat Net : annonces, offres et cycles d'attribution.

- La planification multi-agents : les agents planificateurs ont la responsabilité de la répartition des tâches.

- Les structures organisationnelles : certains agents ont des responsabilités fixes pour des tâches particulières.

*3.3.1.3. La négociation*

La négociation intervient lorsque des agents interagissent pour prendre des décisions communes, alors qu'ils poursuivent des buts différents.

Les trois principales voies de recherche sur la négociation sont :

- Les langages de négociation : il s'agit d'étudier les primitives de communication pour la négociation, leur sémantique et leur usage dans les protocoles.

- Le processus de négociation : il s'agit de proposer des modèles généraux de comportements des agents en situation de négociation.

- Le processus décisionnel : il s'agit d'étudier les algorithmes de comparaison des sujets de négociation, les fonctions d'utilité, et les caractéristiques des préférences des agents (positions, concessions et critères de compromis).

De nombreuses techniques de négociation ont été proposées. Elles sont, soit centrées sur l'environnement, soit centrées sur les agents.

L'idée de la négociation centrée sur l'environnement est de voir comment on peut agir sur l'environnement, en décrivant les règles qui régissent son fonctionnement, pour faciliter le bon fonctionnement des agents dans la résolution de conflit via la négociation.

Pour la négociation centrée sur l'agent, le problème n'est plus d'adapter le contexte à la négociation, mais le comportement de l'agent compte-tenu des propriétés du contexte donné.

### 3.4. Application des SMA dans les télécommunications

Les recherches dans le domaine des télécommunications sont guidées par les services et s'orientent vers :

- Le développement des middlewares qui représentent des architectures logicielles intégrées et interfacées au réseau.

- Le développement de la convergence entre fixe, mobile, son et données [BEN 03].

- La réduction du temps de développement de nouveaux services [JRA 03].

Les objectifs de ces recherches sont :

- Le développement de services en réponse aux offres concurrentes.

- Etendre le rôle traditionnel des opérateurs de télécommunications [BOU 92].

Le tableau suivant montre les différents domaines des télécommunications couverts par les entités agent.

| Domaines | Applications |
|---|---|
| Web et Internet | Assistant pour l'organisation de voyages |
|  | Médiation dans le commerce électronique |
|  | Création de communautés *web* |
| Services et réseaux de télécommunications | Réseaux privés virtuels (VPN) et entreprises virtuelles |
|  | Mobile de $3^{ème}$ génération et réseau intelligent (RI) |
|  | Supervision et gestion de réseaux |
| Ingénierie logicielle des télécommunications | Standardisation d'interopérabilité agent |
|  | Méthodologie et atelier logiciel agent |

**Tableau 3.1**. *Domaines des télécommunications couverts par les entités agent*

#### 3.4.1. Applications des SMA dans le web

Parmi les applications des SMA dans le Web, on s'intéresse aux expérimentations réalisées dans le domaine des services d'assistance dans l'organisation et le déroulement de voyages.

Ces expérimentations ont pour but d'intégrer les différents moyens de notification tels que le fax, et l'e-mail et d'offrir des services complémentaires prenant en compte le profil de l'utilisateur, tels que le calcul d'un itinéraire et la réservation de billets de transport.

On peut généraliser l'utilisation des agents dans le web au domaine du commerce électronique, l'intérêt des SMA dans ce contexte est très lié à l'évolution du commerce électronique vers un marché ouvert très libéral.

L'utilisation des agents dans ce domaine permet d'optimiser la recherche de produit, la comparaison des offres et la négociation des termes d'achat.

Une autre application des agents dans le Web consiste à créer des communautés virtuelles pour faciliter l'échange d'informations entre les cybernautes. Les recherches dans ce contexte s'intéressent à la création de communautés d'agents et à la migration des agents entre communautés [BOU 02].

### *3.4.2. Application des SMA aux réseaux privés virtuels*

Les réseaux privés virtuels (VPN) sont des connexions sécurisées reliant deux réseaux privés via un réseau public, ils permettent d'utiliser ces réseaux publics pour étendre le concept d'Intranet au-delà du réseau privé d'une organisation et ce en préservant la sécurité de la communication.

Le concept de base sur lequel repose les VPNs est le TUNNELING ou ENCAPSULATION de paquets de la couche IP ou de la couche réseau (suivant le protocole utilisé).

Les systèmes multi-agents sont utilisés pour automatiser la négociation des ressources réseau dans ce contexte.

L'intérêt de l'automatisation de la négociation entre les différents acteurs est étroitement lié au nombre d'acteurs présents sur le marché et à la dissociation des opérateurs réseau et des fournisseurs de services de télécommunications [BOU 02].

*3.4.3. Utilisation des SMA dans le cadre des mobiles de troisième génération et du RI*

Les caractéristiques des mobiles de troisième génération sont définies, en Europe, par la norme UMTS (Universal Mobile Telecommunication System). Cette norme introduit les infrastructures VHE (Virtual Home Environment) et PCS (Personnel Communication Support).

L'infrastructure VHE représente un middleware qui masque à l'utilisateur d'un mobile lors du changement de domaine les capacités réelles du réseau.

L'infrastructure PCS définit des polices personnalisées pour traiter les appels en fonction de critères variés tels que l'heure de réception, l'émetteur et le contenu.

Pour répondre à la convergence des réseaux mobiles et de l'Internet et à l'usage international des mobiles, les évolutions de réseaux surtout au niveau de l'administration et du contrôle de service doivent être considérées. Les systèmes multi-agents apportent des éléments de solution à ces évolutions, mais la plupart des travaux de recherche sur l'application des SMA aux réseaux mobiles sont récents et ont été peu expérimentés.

Dans ce contexte, un agent est considéré comme un composant dont les interactions avec les autres composants portent sur la fourniture et sur la qualité de service. C'est un moyen pour uniformiser le traitement des composants de services et des profils utilisateurs [BOU 02] [SAM 05].

*3.4.4. Application des SMA à la supervision et gestion de réseaux*

La supervision de réseau, notamment la reconnaissance de pannes à partir d'observations locales d'alarmes peut être considérée comme une résolution distribuée de problèmes [PLU 96].

Dans ce domaine d'application, les agents représentent des composants logiciels supervisant de façon décentralisée des ressources réseau. Ils sont utilisés pour développer des stratégies de contrôle de la surcharge du réseau, ainsi que pour développer des stratégies de coopération permettant de coordonner la supervision de ressources dépendant de différentes autorités.

Les travaux sur l'application des agents à la gestion de réseau sont encore peu développés, mais il existe des solutions alternatives fondées sur des approches distribuées pour traiter de problème dans lesquels l'organisation des agents est hiérarchique et où le contexte n'est pas un environnement ouvert. Ces solutions ne sont pas basées sur les langages agents mais sur les mécanismes de reconnaissance de scénarios [BOU 02].

**3.5. Conclusion**

Les systèmes muti-agents constituent un thème de recherche en pleine évolution. Ils font intervenir plusieurs domaines de recherche tels que l'intelligence artificielle, les systèmes répartis, la psychologie cognitive et la biologie.

Dans ce chapitre, nous avons présenté l'approche multi-agents, tout en essayant de clarifier la terminologie du domaine comme IAD, SMA.

Les principales conclusions que nous pouvons faire sont les suivantes :

Un système multi-agents s'adapte mieux à la réalité des environnements complexes que l'intelligence artificielle classique.

Un système multi-agents enrichit le processus de résolution de problèmes en le partageant entre plusieurs agents.

L'évolution des machines parallèles est un atout pour l'approche SMA.

# Chapitre 4

# Accès dynamique au spectre

## 4.1. Introduction

Les techniques d'intelligence artificielle pour l'apprentissage et la prise de décision peuvent être appliquées à la conception de systèmes efficaces de la radio cognitive. Le concept de l'apprentissage automatique peut être appliqué à la radio cognitive pour la maximisation des capacités d'accès au spectre dynamique. L'architecture du système proposé est illustrée à la figure 4.1.ci-dessous.

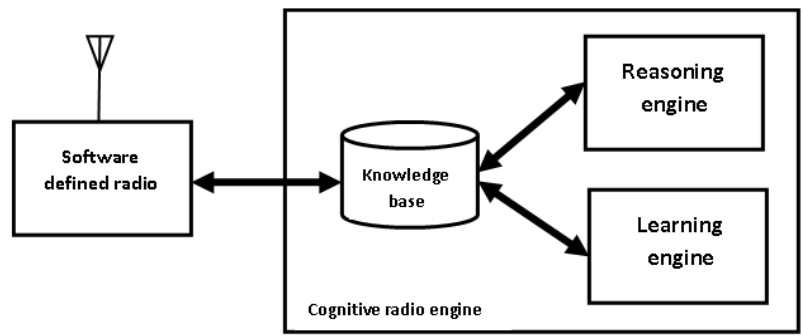

**Figure 4.1.** *Architecture de la radio cognitive avec l'apprentissage automatique*

Ici, la base de connaissances (knowledge base) maintient les états du système et les actions disponibles. Le moteur de raisonnement (Reasoning engine) utilise la base de connaissances pour choisir la meilleure action. Le moteur d'apprentissage (Learning engine) effectue la manipulation des connaissances basées sur l'information observée (par exemple des informations sur la disponibilité des canaux, le taux d'erreur dans le canal). Dans la base de connaissances, deux structures de données, à savoir, le prédicat et l'action, sont définis.

Le prédicat (règle d'inférence) est utilisé pour représenter l'état de l'environnement. Sur la base de cet état, une action peut être effectuée pour modifier l'état de telle sorte que les objectifs du système peuvent être réalisés. Par exemple, un prédicat peut être définie comme la «modulation==QPSK AND SNR == 5dB », tandis que l'action peut être définie comme «mode de modulation en baisse» avec pré-condition « SNR $\leq$ 8dB » et post-condition «modulation ==BPSK ».

Compte tenu de l'entrée (qui est obtenue à partir de la mesure), le moteur de raisonnement correspond à l'état actuel (modulation et SNR dans ce cas) avec les prédicats et détermine les résultats sous-jacents (vrai ou faux). Puis, à partir de l'ensemble des résultats des prédicats, une action appropriée est prise.

Dans l'exemple ci-dessus, si le SNR actuel est égal à 5 dB et la modulation QPSK est en cours, la pré-condition sera vraie et le prédicat sera actif. En conséquence, le moteur cognitif va décider de réduire le mode de modulation. Dans ce cas, la modulation sera modifiée pour BPSK, comme indiqué dans la post-condition correspondante.

Un algorithme d'apprentissage est utilisé pour mettre à jour à la fois l'état du système et les mesures disponibles en fonction de l'environnement radio. Cette mise à jour peut être faite en utilisant une fonction objectif (par exemple, réduire le taux d'erreur binaire) avec un objectif de déterminer la meilleure action compte tenu de l'entrée (par exemple la qualité du canal) et les connaissances disponibles. Différents algorithmes d'apprentissage peuvent être utilisés dans un réseau radio cognitif (le modèle de Markov caché, les réseaux de neurones, ou les algorithmes génétiques) [CLA 07].

**4.2. Algorithmes intelligents**

Les radios cognitives doivent avoir la capacité d'apprendre et d'adapter leur transmission sans fil selon l'environnement radio ambiant. Les algorithmes intelligents tels que ceux basés sur l'apprentissage automatique, les algorithmes génétiques, et la logique floue sont donc essentiels pour la mise en œuvre de la technologie de la radio cognitive. En général, ces algorithmes sont utilisés pour observer l'état de l'environnement sans fil et de construire des connaissances sur l'environnement.

Cette connaissance est utilisée par une radio cognitive pour adapter sa décision sur le spectre d'accès. Par exemple, une radio cognitive (un utilisateur secondaire) peut observer l'activité de transmission de l'utilisateur primaire sur des canaux différents. Cela permet à la radio cognitive de développer les connaissances sur l'activité des utilisateurs primaires sur chaque canal. Cette connaissance est ensuite utilisée par la radio cognitive pour décider quelle voie d'accès choisir afin que les objectifs de performance souhaités peuvent être atteints (par exemple, le débit est maximisé alors que l'interférence ou les collisions causées aux utilisateurs primaires sont maintenues en dessous du niveau cible).

*4.2.1. Réseaux de neurones*

Les réseaux de neurones artificiels sont constitués de neurones artificiels interconnectés les uns avec les autres pour former une structure qui reproduit les comportements des neurones biologiques. Ils peuvent être utilisés dans n'importe quelle phase du cycle de la cognition de la RC [KAT 10].

Le réseau de neurones fournit un modèle à boite noire pour la relation non linéaire entre les entrées (paramètres réseau par exemple) et les sorties (les performances du réseau par exemple). Ce modèle de réseau de neurones peut apprendre à partir des données d'apprentissage qui peuvent être obtenues d'une manière en ligne lorsque les données de mesure en temps réel sont disponibles. Bien que l'apprentissage d'un modèle de réseau de neurones nécessite une grande quantité de ressources de calcul, le calcul de la sortie est beaucoup plus simple et il encourt seulement un léger surcoût.

Par conséquent, ce modèle est approprié pour un réseau radio cognitif pour laquelle une réponse rapide à l'évolution de l'environnement radio est exigée d'un utilisateur secondaire. Par exemple, l'utilisateur secondaire doit interrompre la transmission dès que l'activité de l'utilisateur primaire sur le même canal est détectée.

Le modèle de réseau de neurones illustré dans la figure 4.2 est composé de couches cachées et une couche de sortie.

Les entrées du modèle sont la qualité du canal, le nombre de trames reçues avec succès, le nombre de trames erronées, et la fraction du temps dans laquelle un canal est détecté.

Les sorties du modèle sont le débit, le délai, et la fiabilité du réseau.

Lors de l'apprentissage de ce modèle de réseau de neurones, toutes les entrées mesurées sont utilisées pour ajuster le poids et pour minimiser l'erreur par rapport aux sorties connues. Cet ajustement est répété jusqu'à ce que l'erreur soit inférieure à un certain seuil. Pour ajuster le poids, l'algorithme de rétro-propagation est utilisé. Les données d'apprentissage comprennent des paramètres réseau et les mesures de performance correspondantes.

Ce modèle de réseau de neurones a été évalué et il a été montré que les performances estimées obtenues à partir de ce modèle de réseau de neurones sont proches de ceux des résultats de la simulation [HOS 09].

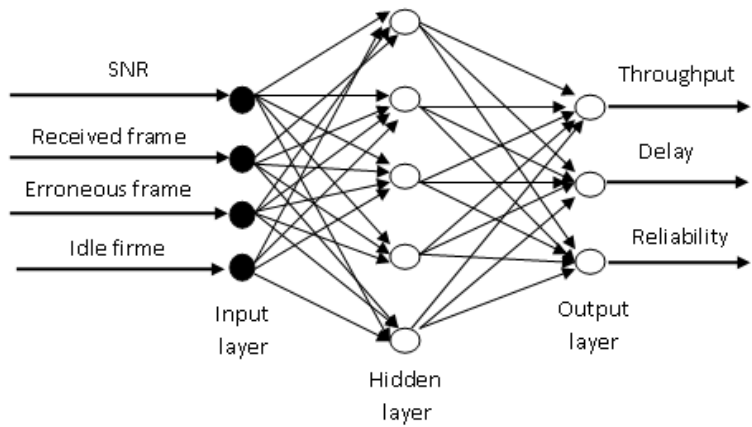

**Figure 4.2.** Un réseau de neurones pour la modélisation des performances de l'IEEE 802.11

Différentes méthodes d'apprentissage automatique appliquées dans les systèmes cognitifs ont été étudiées [BAN 12]. Parmi ces méthodes, les réseaux de neurones qui ont été adoptés dans la détection du spectre et pour l'adaptation des paramètres radio dans les RC [FEH 05].

Des programmes d'apprentissage ont été proposés, ils sont basés sur des réseaux neuronaux artificiels où le taux de données est étudié par rapport à la qualité de la liaison et la force du signal du terminal sans fil [KAT 10].

La prédiction de la performance et la capacité du réseau radio cognitive peut être faite en utilisant les réseaux de neurones multicouches [BAL 09] ou bien en utilisant les cartes de Kohonen [DEM 09].

### *4.2.2. Logique floue*

La logique floue fournit un moyen simple d'obtenir la solution à un problème basé sur l'information imprécise, bruyante, et inachevée.

Au lieu d'employer la formulation mathématique compliquée, la logique floue emploie un ensemble flou de fonctions d'adhésion et de règles d'inférence pour obtenir la solution qui satisfait les objectifs désirés.

Généralement, il y a trois composants importants dans un système de contrôle de logique floue : fuzzifier, processeur de logique floue et le defuzzifier. Tandis que le fuzzifier est employé pour tracer les entrées réelles en les rendant floues, le processeur de logique floue met en application un moteur d'inférence pour obtenir la solution basée sur les ensembles de règles prédéfinies. Alors que le defuzzifier est appliqué pour transformer la solution à la production réelle. La figure ci-dessous montre un schéma synoptique d'un contrôleur flou.

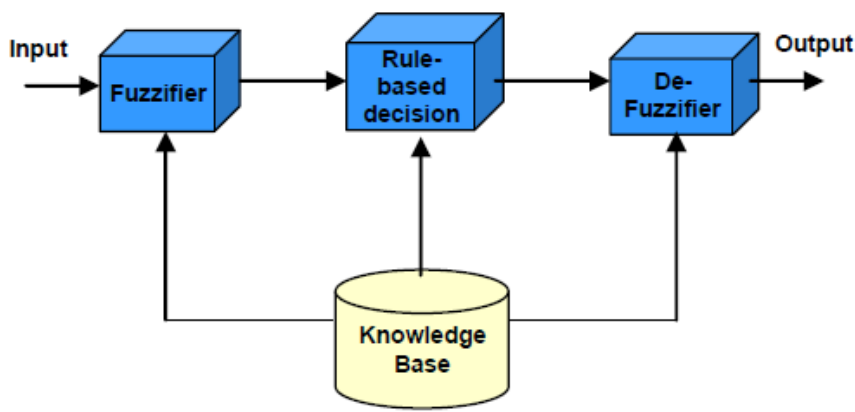

**Figure 4.3.** Un contrôleur flou

De nos jours, les techniques de l'Intelligence Artificielle sont très utilisés pour résoudre quelques problèmes dans le domaine des Télécommunications. Par exemple la logique floue a été proposé comme solution au problème du handover dans les réseaux cellulaires par les auteurs de [GIU 08b] [MAT 00] [HOU 06].

Dans la littérature, la logique floue est souvent utilisée lorsqu'il s'agit de cross-layer optimisation [SHA 12] [YAN 07] dans les systèmes de radio cognitive. Cette technique est utilisée aussi pour la sélection de canaux [FUQ 08] ou sélection du SU le plus approprié pour accéder au spectre en tenant compte de l'efficacité, la mobilité et la distance par rapport au PU [HST 08].

L'utilisation de la logique floue dans la détection coopérative du spectre peut fournir une flexibilité supplémentaire aux méthodes de combinaisons existantes [MAT 09].

La logique floue est souvent combinée avec les réseaux de neurones dans les réseaux radio cognitive [KRL 03] [GIU 08a]. Elle est aussi utilisée comme approche pour le routage multi sauts [MAS 11] ou bien pour la détection des utilisateurs malveillants dans les réseaux radio cognitive [ABO 11].

*4.2.3. Algorithmes génétiques*

Les algorithmes génétiques (AG) appartiennent au domaine de l'intelligence artificielle. Un algorithme génétique est une technique heuristique de recherche inspirée biologiquement qui imite le processus de l'évolution naturelle.

Les AG sont généralement utilisées pour développer un modèle de moteur cognitif [KAT 10], ils sont également utilisés dans la RC pour résoudre des problèmes d'optimisation et pour configurer les paramètres RC lorsqu'il y a un changement dans l'environnement sans fil [RON 04] [NEW 07].

Les AG sont appropriées pour les problèmes liés à la RC [RIE 04], car ils offrent une quantité importante de puissance et de flexibilité étant donné que les RC sont susceptibles de faire face à des environnements dynamiques et des mises à niveau de radio en raison de l'avancement de la technologie.

Les AG sont utilisés pour la RC pour proposer un gestionnaire de ressources cognitives pour sélectionner un algorithme à partir d'une boîte à outils pour résoudre les problèmes.

**4.3. Accès dynamique au spectre**

La croissance explosive des services sans fil ces dernières années illustre la demande croissante des consommateurs pour les communications, ainsi le spectre devient plus encombré. Nous savons que l'allocation du spectre statique est un problème majeur dans les réseaux sans fil. Généralement, ces allocations conduisent à une utilisation inefficace du spectre et elles créent des trous ou des espaces blancs dans le spectre. Pour résoudre le problème de l'encombrement, les réseaux RC utilisent l'accès dynamique au spectre.

La communication coopérative est connue comme un moyen pour surmonter les limites des systèmes sans fil [LET 07]. Cependant, puisque les utilisateurs ont généralement une connaissance limitée de leur environnement, nous prétendons que le comportement coopératif peut leur fournir les informations nécessaires pour résoudre les problèmes globaux.

Fondamentalement, un utilisateur secondaire ne possède pas une licence pour son utilisation du spectre et il peut y accéder soit de manière opportuniste soit par la coexistence avec les utilisateurs voisins autorisés. Ce type d'accès est appelé « partage de licence » et un assez grand nombre de solutions existent déjà dans la littérature [YAN 10] [NIY 08] [MIR 10b].

Nous avons trouvé un grand nombre de travaux proposés relatifs à l'accès au spectre, ceux utilisant les ventes aux enchères, un grand nombre d'approches utilisent la théorie des jeux, mais les approches utilisant les chaines de Markov sont moins nombreuses. Cependant, quelques recherches ont été faites dans ce domaine en utilisant les systèmes multi-agents.

*4.3.1. Accès au spectre en utilisant les Enchères*

Les enchères sont basées sur le concept de vente et d'achat des biens ou de services. Le but principal de l'utilisation des enchères dans les réseaux RC est de fournir une motivation aux utilisateurs secondaires pour maximiser leur utilisation du spectre. Afin d'utiliser pleinement le spectre, l'allocation dynamique du spectre utilisant les enchères est devenue une approche prometteuse qui permet aux utilisateurs secondaires de louer des bandes inutilisées par les utilisateurs primaires.

Dans les solutions basées sur les enchères, chaque canal est assigné à un seul réseau, c'est à dire qu'il n'y a pas la notion de SU et de PU dans le même canal. Dans la littérature, deux possibilités s'offrent:
- Soit le régulateur alloue les canaux aux utilisateurs primaires, ces derniers allouent indépendamment les portions inutilisées de leur canal aux SU [HUA 04].

- Soit le régulateur alloue le droit d'être SU ou PU dans le canal.

Une plateforme pour l'allocation du spectre dans la RC est proposée utilisant les enchères où le régulateur peut allouer simultanément les droits d'être PU ou SU dans le canal [MIR 11a].

Une autre façon d'utiliser les ventes aux enchères est proposée où les auteurs ont prouvé que dans certains scénarios le spectre est utilisé efficacement lorsque plusieurs SU gagnent l'accès à un seul canal, c'est ce qui distingue leur méthode avec les enchères traditionnelles où un seul utilisateur peut gagner [YON 08].

Dans ces solutions, les comportements des utilisateurs sont mensongers, de sorte que le gestionnaire centralisé ne peut pas optimiser la fonction d'utilité globale du réseau [MIR 11b].

### *4.3.2. Accès au spectre en utilisant la Théorie des jeux*

La Théorie des jeux peut être définie comme un cadre mathématique qui se compose de modèles et de techniques utilisés pour analyser le comportement itératif des individus préoccupés par leur propre bénéfice. Ces jeux sont généralement divisés en deux types [ELN 08] : jeux coopératifs et jeux compétitifs.
**- Jeux coopératifs**: tous les joueurs sont préoccupés par tous les gains globaux et ils ne sont pas très inquiets de leur gain personnel. Certains travaux récents [YAN 10] [ZHA 09] utilisent la théorie des jeux coopératifs pour réduire la puissance de transmission des utilisateurs secondaires afin d'éviter de générer des interférences avec les transmissions des utilisateurs primaires.

- Jeux compétitifs: chaque utilisateur est principalement préoccupé par son gain personnel et donc toutes ses décisions sont prises de manière compétitive et égoïste. Dans la littérature existante, nous avons constaté que les concepts théoriques du jeu ont été largement utilisés pour des attributions de fréquences dans les réseaux RC [NIY O8] [WAN 10] [TAN 07], où lorsque les utilisateurs primaires et secondaires participent à un jeu, ils ont un comportement rationnel pour choisir les stratégies qui maximisent leurs propres gains.

La propriété la plus connue des approches de la théorie des jeux est appelée équilibre de Nash (EN). Dans l'EN, chaque joueur est supposé connaître les stratégies d'équilibre des autres joueurs, et aucun joueur n'a rien à gagner en changeant sa propre stratégie.

Certains des travaux existants utilisant la théorie des jeux pour l'accès dynamique au spectre sont mentionnés ici. Par exemple, pour Xiuli [XIU 07], les utilisateurs primaires sont conscients de l'existence des utilisateurs secondaires et ils ont une priorité plus élevée sur l'accès au spectre. Les utilisateurs primaires adoptent les rôles de leaders en sélectionnant un sous-ensemble d'utilisateurs secondaires et leur garantit un accès au spectre. Les utilisateurs secondaires augmentent leurs valeurs d'utilité en termes d'accès au spectre et payent les utilisateurs primaires. Alors que pour Lai [LAI 08], les utilisateurs primaires n'ont pas les connaissances sur leur environnement et voisinage, donc ils ne sont pas conscients de la présence des utilisateurs secondaires, ces derniers sont autorisés seulement à accéder au spectre de façon opportuniste.

Un jeu intéressant est proposé par Yun [YUN 09] où l'utilisateur primaire détermine le prix de spectre en premier en se basant sur la qualité du spectre, ensuite, l'utilisateur secondaire décide de la quantité de spectre qu'il doit acheter en observant les prix.

Dans les jeux de négociation, les joueurs individuels ont la possibilité de coopérer afin de parvenir à un accord mutuel. En même temps, ces joueurs peuvent avoir des conflits d'intérêt et aucun accord ne peut être fait avec n'importe quel joueur individuel sans son approbation. Pour les réseaux RC, les jeux de négociation sont appliqués pour allouer des bandes de fréquences dans les réseaux centralisés et décentralisés [DEB 08].

Il faut mentionner que même si les jeux coopératifs et compétitifs ne s'intéressent qu'à la résolution de l'EN et l'analyse de ses propriétés, ils ne fournissent pas de détails sur l'interaction des joueurs pour atteindre cet équilibre [WAN 10].

*4.3.3. Accès au spectre en utilisant les approches de Markov*

Les approches de la théorie des jeux ne modélisent pas l'interaction entre les utilisateurs secondaires et primaires pour l'accès au spectre. Cette modélisation peut être réalisée en utilisant efficacement les chaînes de Markov [MIR 11b].

Peu de recherches ont été effectuées dans ce domaine, par exemple, Xin [XIN 10] propose un modèle de Markov, où chaque utilisateur secondaire sélectionne au hasard sa propre chaîne plutôt que d'échanger des messages de contrôle avec les utilisateurs secondaires voisins. Sinon, il y a aussi l'utilisation du modèle de Markov pour prédire les comportements du canal [GEI 07].

Certains auteurs utilisent le CTMC (Continious Time Markov Chains) [AHM 09] [ZHA 08] pour capturer l'interaction entre les utilisateurs primaires et secondaires. Les deux modèles avec et sans files d'attente sont analysés et la dégradation du débit à cause des interférences des utilisateurs secondaires est compensée. Les modèles CTMC obtiennent de bonnes statistiques entre les statistiques d'équité et d'efficacité.

Néanmoins, un nombre limité de travaux ont utilisé les chaînes de Markov pour l'accès au spectre sans licence. Parmi ces derniers, un modèle est proposé par Xing [XIN 06] pour atteindre des assignations équitables de fréquences entre les utilisateurs secondaires. Cette approche vise plus spécifiquement sur l'utilisation efficace du spectre, en évitant les interférences.

*4.3.4. Accès au spectre en utilisant les Systèmes Multi Agents*

L'association des SMA avec la RC assure un futur remarquable pour la gestion optimale des fréquences (en comparaison avec les techniques de contrôle rigides proposées par les opérateurs de télécommunications). Dans le cas de l'utilisation des bandes sans licence, le terminal RC doit coordonner et coopérer pour un usage meilleur du spectre sans causer d'interférences.

Ahmed [AHM 11] propose une architecture basée sur les agents où chaque terminal RC est équipé d'un agent intelligent, il y a des modules pour collecter les informations à propos de l'environnement radio et bien sur les informations collectées seront stockées dans une base de connaissance partagée qui sera consultée par tous les agents. L'approche proposée est basée sur les SMA coopératifs (les agents ont des intérêts en commun). Ils collaborent en partageant leurs connaissances pour augmenter leur gain individuel ainsi que collectif.

Des agents sont déployés sur les terminaux RC des utilisateurs primaires PU et des utilisateurs secondaires SU et coopèrent entre eux dans les travaux proposés par Mir [MIR 10a] [MIR 10b] et Haykin [HAY 05]. Par SMA coopératif, on veut dire que les agents PU échangent des t-uples de messages dans le but de s'améliorer eux-mêmes ainsi que le voisinage des agents SU. Ils proposent que les SU doivent prendre leur décision en se basant sur la quantité du spectre disponible, le temps et le prix proposé par les agents PU. Et ils doivent commencer le partage du spectre dès qu'ils trouvent une offre appropriée (Sans attendre la réponse de tous les PU). En d'autres termes, l'agent SU doit envoyer des messages à l'agent PU voisin approprié, et bien sur le PU concerné doit répondre à ces agents pour faire un accord sur le partage du spectre. Et bien sur après la fin de l'utilisation du spectre, le SU doit payer le PU.

Une comparaison est faite par Mir [MIR 10b] entre un agent et une RC. Principalement, les deux sont conscients de leur environnement à travers les interactions, détection, surveillance. Ils sont autonomes, ils peuvent résoudre des tâches en se basant sur leurs propres capacités et bien sur ils peuvent coopérer avec leurs voisins en échangeant des informations.

| Agent | Radio cognitive |
|---|---|
| Conscience de l'environnement à travers les anciennes observations | Détecte les espaces blancs du spectre et les signaux des utilisateurs primaires |
| Agit à travers des actionneurs | Décider quelle bandes/canaux vont être sélectionnés |
| Interaction via la coopération | Interaction via le balisage |
| Autonomie | Autonomie |
| Travaillent ensemble pour atteindre des objectifs communs | Travaillent ensemble pour un partage efficace du spectre |
| Contient une base de connaissance avec des informations locales et sur les agents voisins | Maintient certains modèles d'utilisation du spectre des utilisateurs primaires voisins. |

**Tableau 4.1.** *Comparaison entre un agent et une Radio Cognitive*

Pour rendre les systèmes de RC pratiques, il faut que plusieurs réseaux RC coexistent entre eux. Cependant, ceci peut générer des interférences. Ben Letaief [LET 07] pense que pour remédier à ce problème, les SU peuvent coopérer pour détecter le spectre aussi bien que pour le partager sans causer d'interférences pour le PU. Pour cela, ils proposent des schémas pour protéger les PU des interférences en contrôlant la puissance de transmission du terminal cognitif.

Mir [MIR 11b] [MIR 10c] propose une coopération entre les PU et les SU et entre les SU seulement. Des agents sont déployés sur les terminaux des utilisateurs pour coopérer et aboutir à des contrats régissant l'allocation du spectre. Les agents SU coexistent et coopèrent avec les agents PU dans un environnement RC Ad-hoc en utilisant des messages et des mécanismes de prise de décision. Vu que les comportements internes des agents sont coopératifs et désintéressés, ça leur permet de maximiser la fonction d'utilité des autres agents sans ajouter de coût conséquent en termes de messages échangés.

Cependant, l'allocation des ressources est un enjeu important dans les systèmes de RC. Il peut être fait en effectuant la négociation parmi les utilisateurs secondaires [TIA 10] [HUS 09]. Dans [TIA 10] les auteurs proposent un modèle basé sur les agents pour la négociation du spectre dans un réseau RC. Mais au lieu de négocier le spectre directement entre des PU et des SU, un agent courtier est inclus. Ce qui veut dire que l'équipement du PU ou du SU ne nécessite pas une grande intelligence vu qu'il n'a pas besoin d'effectuer la détection du spectre ou autre chose. L'objectif de cette négociation est de maximiser les bénéfices et les profits des agents pour satisfaire le SU. Les auteurs ont proposé deux situations, la première utilise un seul agent qui va exploiter et dominer le réseau, et dans la deuxième, il va y avoir plusieurs agents en concurrence.

Jiang [JIA 07] a étudié la RC dans les réseaux WLAN et la possibilité d'introduire la technologie d'agents, en d'autres termes ils essayent de résoudre le problème de l'allocation des ressources radio en associant la gestion des ressources WLAN dans un environnement décentralisé, ceci en utilisant les SMA. Pour cela, ils proposent une approche basée sur les SMA pour le partage d'information et la distribution des décisions parmi de multiples WLANs d'une manière distribuée.

Les interférences causées par l'acquisition des canaux dans un système cellulaire au cours des Handovers peuvent être réduites selon [RAI 08] en utilisant une RC pour la gestion du Handover. En effet, la mobilité du terminal lui impose un comportement différent au moment du changement de zones. Le terminal doit assurer la continuité de service de ses applications ainsi que la gestion efficace du spectre. Les auteurs proposent une approche qui utilise la négociation, l'apprentissage, le raisonnement et la prédiction pour connaitre les besoins des nouveaux services dans les réseaux sans fil modernes. Ils proposent un algorithme à exécuter par le terminal cognitif mobile lors de la phase de Handover.

Le SMA contient plusieurs agents intelligents en interaction entre eux. Chaque agent peut faire la détection et l'apprentissage. L'agent peut sélectionner les comportements basés sur l'information locale et tenter de maximiser les performances globales du système. Cheng [CHE 10] a décrit une nouvelle approche basée sur l'apprentissage par renforcement multi-agent qui est utilisée sur des réseaux RC ad-hoc avec contrôle décentralisé. En d'autres termes, ils ont mis en place plusieurs scénarios de RC et ils affectent à chaque cas une récompense ou une pénalité. Les résultats de cette approche ont montré qu'avec cette méthode, le réseau peut converger à un partage équitable du spectre et bien sur elle permet de réduire les interférences avec les utilisateurs primaires PU.

Une approche très intéressante est proposée par Yau [YAU 10] où les auteurs ont appliqué l'apprentissage par renforcement sur des cas uni-agent SARL et multi-agent MARL pour atteindre la sensibilité et l'intelligence. Ils montrent dans leurs résultats que les SARL et les MARL réalisent une action commune qui donne un meilleur rendement à l'échelle du réseau. Ils ont fini par dire l'apprentissage par renforcement RL est un algorithme adapté pour être appliquée dans la plupart des schémas d'application.

Dans la solution proposée par Kok-Lim [YAU 11], un mécanisme d'apprentissage local comme le MARL est disponible pour chaque agent. L'apprentissage local fournit pour chaque agent une récompense pour qu'il puisse prendre la bonne décision et choisir la meilleure action. Ils ont modélisé chaque nœud de communication de SU comme un agent d'apprentissage car le récepteur et l'émetteur partagent une issue commune d'apprentissage ou de connaissance.

Les auteurs ont présenté le LCPP (Locally Confined Payoff Propagation) qui est une fonction importante dans l'apprentissage par renforcement dans les SMA pour atteindre l'optimalité dans la coopération entre agents dans un réseau RC distribué.

Un schéma de sélection de canal sans négociation est considéré pour le multi utilisateur et le multi-canal [HUS 09]. Afin d'éviter les collisions encourus par la non-coordination, chaque SU apprend à sélectionner les canaux en fonction de ses expériences. L'apprentissage par renforcement multi-agents est appliqué dans le cadre du Q-learning en considérant les utilisateurs secondaires comme une partie de l'environnement. Dans un tel schéma, chaque utilisateur secondaire détecte des canaux et ensuite choisit un canal de fréquence ralenti à transmettre les données, comme si aucun autre utilisateur secondaire n'existe. Si deux SU choisissent le même canal de transmission de données, ils vont entrer en collision les uns avec les autres et les paquets de données ne peuvent pas être décodés par le récepteur. Cependant, les utilisateurs secondaires peuvent essayer d'apprendre comment s'éviter les uns les autres.

Galindo-Serrano [GAL 09] s'est intéressé à l'utilisation de l'IEEE 802.22, et ils ont proposé un algorithme nommé "Decentralized Q-learning" basé sur la théorie de l'apprentissage multi-agent pour faire face au problème des interférences causées aux PU. Ils ont modélisé le réseau secondaire à l'aide de SMA où les différents agents sont des Stations de Base secondaires de l'IEEE 802.22 WRAN. Ils ont prouvé que le SMA proposé est capable d'apprendre automatiquement la politique optimale pour maintenir la protection pour les PU contre les interférences.

Jiandong [JIA 10] et Mir [MIR 11a] ont utilisé les SMA pour concevoir un nouveau cycle de cognition avec les relations complexes d'interaction entres les différents PU, SU et les environnements sans fil coexistant et les chaines de Markov cachées pour modéliser les interactions entre les utilisateurs et l'environnement. Les résultats de cette approche ont montré que l'algorithme peut garantir l'équité entre les utilisateurs.

Ce qui pourrait rendre l'utilisation des SMA dans la RC intéressante et plus concrète, c'est l'existence d'une plateforme de simulation pour tester les travaux proposés. C'est ce que justement proposent les auteurs de [DZI 09]. Leur plateforme permet d'étudier l'aspect émergent et comportemental des réseaux RC hétérogènes.

Amraoui [AMR 12a] [AMR 12b] et Benmammar [BEN 12] s'intéressent à améliorer la fiabilité du lien sans fil afin de garantir une bonne qualité de service aux terminaux RC mobiles en intégrant les systèmes multi agents.

**4.4 Conclusion**

Dans ce chapitre, nous avons présenté diverses méthodes d'accès au spectre en partant des enchères où la fonction utilité du réseau n'est pas optimisé à chaque fois car elle dépend du comportement des utilisateurs, et passant ensuite par la théorie des jeux qui est largement utilisée dans ce domaine car elle permet d'aboutir à un équilibre entre les utilisateurs qui assure une gestion efficace du spectre.

Nous avons ensuite cité quelques travaux réalisés à l'aide des modèles de Markov qui en plus des méthodes précédentes fournit une modélisation de l'interaction entre les utilisateurs secondaires et primaires. Et pour finir, nous nous sommes concentrés sur l'utilisation des systèmes multi agents dans l'accès dynamique au spectre, cependant cette méthode a été exploité par une minorité de chercheurs (en comparaison avec la théorie des jeux) pour résoudre le problème de l'allocation du spectre.

Différentes approches utilisant les SMA dans la RC sont étudiées, celles proposant une coopération entre les utilisateurs secondaires seulement, d'autres proposant une coopération entre les utilisateurs primaires et secondaires et celles proposant d'intégrer un agent courtier pour négocier le spectre, sachant que la plupart des travaux utilisent l'apprentissage par renforcement.

**Bibliographie**

# Index

# Liste de livres conseillés pour approfondissement